\documentclass[12pt]{article}
\usepackage{amsthm}
\usepackage{eurosym}
\usepackage{amsfonts}
\usepackage{amssymb,amsmath}
\usepackage[dvips]{color}
\usepackage{amscd}
\usepackage{tikz}
\usepackage{mathtools}
\usepackage{graphicx}
\usepackage{caption}
\usepackage{subcaption}
\usepackage[all,cmtip]{xy}

\newcommand{\xdownarrow}[1]{ {\left\downarrow\vbox to #1{}\right.\kern-\nulldelimiterspace} }
\usetikzlibrary{decorations.pathreplacing}

\usetikzlibrary{patterns}

\def\and{\mathrm{and}}

\newtheorem{prop}{Proposition}

\newcommand{\ee}{\end{equation}}
\newcommand{\bea}{\begin{eqnarray}}
\newcommand{\eea}{\end{eqnarray}}
\newcommand{\beas}{\begin{eqnarray*}}
\newcommand{\eeas}{\end{eqnarray*}}
\newcommand{\ba}{\begin{array}}
\newcommand{\ea}{\end{array}}

\newcommand{\nbox}{{\,\lower0.9pt\vbox{\hrule \hbox{\vrule height 0.2 cm \hskip 0.19 cm \vrule height 0.2 cm}\hrule}\,}}

\def\href#1#2{#2}
\theoremstyle{plain}

\begin{document}

\begin{titlepage}
\hfill
\vbox{
    \halign{#\hfil         \cr
           } 
      }  

\hbox to \hsize{{}\hss \vtop{ \hbox{}

}}

%

\vspace*{20mm}

\begin{center}

{\large \textbf{Entanglement and mixed states of Young tableau states\\ \vspace{0.4cm} in gauge/gravity correspondence }}

{\normalsize \vspace{10mm} }

{\normalsize {Hai Lin${}^{1,2}$}, Yuwei Zhu${}^{1,2}$}

{\normalsize \vspace{10mm} }

{\small \emph{${}^1$\textit{Yau Mathematical Sciences Center, Tsinghua University,
Beijing 100084, China
}} }

{\normalsize \vspace{0.2cm} }

{\small \emph{$^2$\textit{Department of Mathematical Sciences, Tsinghua University,
Beijing 100084, China
\\
}} }

{\normalsize \vspace{0.4cm} }

\end{center}

\begin{abstract}

We use entangled multimode coherent states to produce entangled giant graviton states, in the context of gauge/gravity duality. We make a smeared distribution of the entangled multimode coherent states on the circle, or on the five-sphere, in the higher dimensional view. In gauge/gravity duality, we analyze the superposition of giant graviton states, and the entangled pairs of giant graviton states. We map a class of angular distribution functions to unitary operations on the pairs. We also use Young tableau states to construct cat states and qudit states. Various bipartite quantum states involving Young tableau states are analyzed, including micro-macro entangled states. Mixed states of Young tableau states are generated, by using ensemble mixing using angular distribution functions, and also by going through noisy quantum channels. We then produce mixed entangled pair of giant graviton states, by including interaction with the environment and using noisy quantum channels.

\end{abstract}

\end{titlepage}

\vskip 1cm

\section{Introduction}

\label{sec_introduction}

The gauge/gravity correspondence \cite%
{Maldacena:1997re,Gubser:1998bc,Witten:1998qj} is a remarkable
correspondence between a quantum system without gravity on the boundary and
a quantum theory with gravity in the bulk. The correspondence reveals the
nature of the emergent spacetime \cite%
{Horowitz:2006ct,Rangamani:2016dms,Berenstein:2005aa,Koch:2009gq}. The bulk
emerges dynamically from the quantum mechanical description that lives in
fewer dimensions on the boundary. The boundary system is described by a
quantum field theory or quantum mechanics with a well-defined global time
near the boundary. Hence, the boundary theory has a well-defined Hilbert
space. There is a correspondence between observables of the bulk spacetime
and observables of the boundary.

This correspondence provides a method for working on quantum gravity by
quantum field theory on the boundary of the spacetime. The correspondence
allows us to perform calculations related to string theory and quantum
gravity from working on the quantum field theory side. On the other hand,
the string theory provides the ultraviolet completion of supergravity, and
is hence a ultraviolet-complete quantum gravity.

In the gauge/gravity correspondence, the quantum field theory side of the
duality is an example of a quantum system with many degrees of freedom. In
such a many-body quantum system, quantum correlations and quantum
entanglement are generic \cite{Horodecki:2009zz}. Quantum entanglement are
important resources for quantum information processing. Many tools and
methods in quantum information are also useful for working on quantum
gravity, since quantum gravity is also described by quantum mechanical
rules. We can perform superpositions and quantum operations such as unitary
transformations on these quantum states on the gravity side.

We analyze states which have interesting gravitational properties. There are
states that are of interest both in the quantum gravity side and in quantum
information theory, such as coherent states and their superpositions and
entanglement. Moreover, there are Young tableau states \cite%
{Corley:2001zk,Berenstein:2004kk,Koch:2009gq}, which are also entangled
states. In gauge/string correspondence, there are backreacted geometries
that correspond to highly excited states in the quantum field theory side,
such as the bubbling geometries. The states in the Hilbert space of the
quantum field theory are explicitly mapped to the gravity side. Since they
live in the same Hilbert space, one can perform quantum superpositions and
quantum operations on these states.

On the quantum field theory side, there are interesting coherent states \cite%
{Berenstein:2005aa,Berenstein:2017abm}. These are the superpositions of
multi-trace states. In the context of gauge/gravity correspondence, BPS
coherent states have gravity dual descriptions. They are important
ingredients in the superposition-induced topology change in quantum gravity
\cite{Berenstein:2017abm}. The transition between Young tableau (YT) states
and coherent states involves a topology change in the dual spacetime. The
superposition formula that gives a Young tableau state or a Brauer state by
superposing coherent states, have been computed \cite%
{Berenstein:2017abm,Lin:2017dnz,Lin:2017vfn}. The superposition induced
topology changes in the gravity side have been observed in \cite%
{Berenstein:2017abm,Lin:2017dnz,Simon:2018laf,Lin:2020qao}.

We also focus on Young tableau states. The YT states, which are labeled by
Young tableaux, are linear combinations of multi-trace states. The single
trace states include the descriptions of close strings, and similarly, the
YT states also describe brane states. The YT states are also dual to giant
gravitons on the gravity side. The giant gravitons can be viewed as
polarized from point gravitons, under the influence of the anti-symmetric
form-fields.

These YT states are entangled states in the product space of the multi-trace
Hilbert spaces \cite{Berenstein:2017abm,Lin:2017dnz}, especially when they
contain a big number of boxes. These states have nontrivial entanglement
stored between different multi-trace Hilbert spaces. The entanglement
entropy of these states, entangled in the multi-trace Hilbert space, have
been computed in \cite{Lin:2017dnz,Berenstein:2017abm}. These states can
have different bases or labelings, and different bases can be transformed
into each other, e.g. \cite{Lin:2017dnz,Lin:2017vfn}. The transformation
between multi-traces and coherent states are also computed, e.g. \cite%
{Lin:2017vfn}. For more details on the mathematics of the underlying
symmetric groups and Young tableaux, see e.g. \cite{Fulton,Sagan,Fulton
Harris} and \cite{Ramgoolam:2018ceu}.

Among other things, we analyze bipartite quantum states involving YT states.
We constructed bipartite entangled states involving YT, which describe
entangled pairs of giant gravitons. We also use YT states to construct cat
states and qudit states. The bipartite entangled states are important in
gauge/gravity duality and quantum information. We further produced mixed
states of YT and mixed entangled pair of YT.

The organization of this paper is as follows. In Sec. 2, we use entangled
multimode coherent states to produce entangled giant graviton states. This
involves entanglement from two portions of the internal five-sphere. In Sec.
3, we produce entangled states, which are entangled between Young tableau
states and trace states. In Sec. 4, we generate mixed states of Young
tableau states, by using ensemble mixing and then by using noisy quantum
channels. In Sec. 5, we produce mixed or noisy entangled pair of giant
gravitons, by including interaction with the environment and by using noisy
quantum channels. Finally, we discuss our results and draw some conclusions
in Sec. 6. In Appendix A, we briefly overview multimode coherent states and
Young tableau states for the convenience of readers.

\section{Entangled YT states}

\label{sec 2} \renewcommand{\theequation}{2.\arabic{equation}} %
\setcounter{equation}{0} \renewcommand{\thethm}{2.\arabic{thm}} %
\setcounter{thm}{0} \renewcommand{\theprop}{2.\arabic{prop}} %
\setcounter{prop}{0}

\subsection{Entangled bipartite states and entanglement between two halves
of the five-sphere}

In this section, we consider a class of multimode coherent states and Young
tableau states. More details of these two types of states are in Appendix A.
This class of multimode coherent states was constructed in \cite%
{Berenstein:2017abm}, and analyzed in further details in \cite%
{Lin:2017dnz,Lin:2020qao}. This multi-mode coherent state is%
\begin{equation}
|Coh(\Lambda )\rangle =\prod_{k=1}^{\infty }\exp (\Lambda ^{k}\frac{%
a_{k}^{\dagger }}{k})|0\rangle _{k}  \label{Coh 01}
\end{equation}%
with normalization $||Coh(\Lambda )||=\sqrt{\mathcal{N}(\Lambda )}$ where%
\begin{equation}
\mathcal{N}(\Lambda )=\exp (\sum\limits_{k=1}^{\infty }\frac{|\Lambda |^{2k}%
}{k})=\frac{1}{(1-|\Lambda |^{2})}.
\end{equation}%
The complex parameter is $\Lambda =|\Lambda |e^{i\theta _{0}}$. The shape of
the multimode coherent states is determined by the expectation value of the
chiral field evaluated on the states \cite{Berenstein:2017abm}. The chiral
field describes the dynamics of the boundary between black and white droplet
regions. The shape is a bump along the droplet boundary \cite{Lin:2004nb}
located at $\theta _{0}=\mathrm{\arg }(\Lambda )$ with magnitude $|\Lambda |$%
. The shape of the bump is calculated by the expectation value of the chiral
field $\langle \hat{\phi}(\theta )\rangle _{|\Psi \rangle }$, where $\hat{%
\phi}(\theta )=\sum_{k>0}(a_{k}\exp (-ik\theta )+a_{k}^{\dagger }\exp
(ik\theta ))$ \cite{Berenstein:2017abm}. See also \cite{Lin:2017dnz} for
detailed analysis. Note that $|Coh(\Lambda )\rangle $ is a multimode
coherent state, and not a single mode coherent state. This state has a
gravity dual in terms of a bump on the five-sphere.

The one-row Young tableau state is \cite{Berenstein:2017abm,Lin:2017dnz}%
\begin{equation}
\lvert \Delta _{n}\rangle =\frac{1}{2\pi i}\oint_{\mathcal{C}}\frac{\mathrm{d%
}w}{w}w^{-n}|Coh(w)\rangle ,
\end{equation}%
where $\mathcal{C}$ is a path that encloses $0$. The $\lvert \Delta
_{n}\rangle $ has unit norm, i.e. $\Vert \Delta _{n}\Vert =1.$ It has
property
\begin{equation}
|Coh(w)\rangle =\sum_{n=0}^{\infty }w^{n}\lvert \Delta _{n}\rangle
,~~~\langle \Delta _{n}|Coh(w)\rangle =w^{n}.  \label{inner product}
\end{equation}%
Hence%
\begin{equation}
\lvert \Delta {_{n}}\rangle =\frac{1}{2\pi }|\Lambda |^{-n}\int_{0}^{2\pi
}d\theta e^{-in\theta }|Coh(\Lambda e^{i\theta })\rangle .
\end{equation}%
$\lvert \Delta {_{n}}\rangle $ is a state with $n$ boxes. They describe the
giant gravitons. There are two different physical interpretations of the
states. The giant graviton is a brane wrapping submanifolds of internal
dimensions and the dual giant graviton is a brane wrapping submanifolds of
the AdS space. One interpretation is that, it can be interpreted as a state
of $n$ Kaluza-Klein (KK) gravitons. Another interpretation is that, it can
also be interpreted as a state of a dual giant graviton with $n$ units of
angular momentum, \cite{McGreevy:2000cw,Grisaru:2000zn,Hashimoto:2000zp}.
These giant gravitons travel along the circle of the five-sphere.

Now we use entangled multimode coherent states to produce entangled giant
graviton states, or in other words, entangled YT states. Let $\Psi \in
\mathcal{H}_{A}\otimes \mathcal{H}_{B}$ and let $\Lambda _{1}\Lambda
_{2}=\Lambda $. We consider a generic quantum superposition of
angularly-correlated multimode coherent states, with a generic complex
distribution function $\alpha (\theta )\in \mathbb{C}$. Depending on the
angular correlations of the two multimode coherent states, there are
different types of entangled states between the two. Generally, the
following is a class of bipartite entangled states,
\begin{eqnarray}
\Psi &=&\Psi \lbrack \alpha (\theta )]:=\frac{1}{\sqrt{\mathcal{N}(\Lambda )}%
}\int_{0}^{2\pi }\frac{d\theta }{2\pi }~\alpha (\theta )~|Coh(\Lambda
_{1}e^{i\theta })\rangle |Coh(\Lambda _{2}e^{-i\theta })\rangle  \notag \\
&=&\frac{1}{\sqrt{\mathcal{N}(\Lambda )}}~\sum_{n_{1}=0}^{\infty
}\sum_{n_{2}=0}^{\infty }\int_{0}^{2\pi }\frac{d\theta }{2\pi }\alpha
(\theta )e^{i(n_{1}-n_{2})\theta }\Lambda _{1}^{n_{1}}\Lambda
_{2}^{n_{2}}\lvert \Delta _{n_{1}}\rangle \lvert \Delta _{n_{2}}\rangle .
\label{entangled state_02}
\end{eqnarray}%
Here $\alpha (\theta )$ is a general complex coefficient, i.e. $\alpha
(\theta )\in \mathbb{C}$. The two angles are oppositely correlated. The
angular distribution function plays the role of superposition of the
bipartite states. We have expanded the entangled states in YT basis which is
the giant graviton basis.

Now we consider a special case when $\Lambda _{2}/\Lambda _{1}\in \mathbb{R}%
_{>0}$ for (\ref{entangled state_02}); In this convention of the parameters,
the state (\ref{entangled state_02}) describes entanglement between two
halves of the circle, since the angles are oppositely correlated, and in the
higher dimensional view, between two halves of the five-sphere. We have the
inclusion $i:S^{1}\hookrightarrow S^{5}$ and the projection $%
p:S^{5}\rightarrow S^{1}$. As we show, they are also equivalent to entangled
giant graviton states. The states of bubbling geometries and the states of
giant gravitons are dual to each other. This includes the case when both
sides are entangled states. Hence we can also interpret $A$ and $B$ as two
subsystems corresponding to the two halves of five-spheres. Hence, this
state may also be viewed as an entangled state between excitations on the
two halves of the circle, or on the two halves of the five-sphere in the
higher dimensional view. These entangled configurations have also been
considered in \cite{Simon:2018laf}. Moreover, quantum states of giant
gravitons, e.g. \cite{Berenstein:2005aa},\cite{Grant:2005qc},\cite%
{Mandal:2005wv}--\cite{Carlson:2011hy}, can also be enumerated by states of
multi-dimensional quantum harmonic oscillators.

For $\alpha (\theta )=1,$ since$~\int_{0}^{2\pi }\frac{d\theta }{2\pi }%
~e^{i(n_{1}-n_{2})\theta }=\delta _{n_{1},n_{2}},$
\begin{eqnarray}
\Psi &=&\frac{1}{\sqrt{\mathcal{N}(\Lambda )}}\int_{0}^{2\pi }\frac{d\theta
}{2\pi }~|Coh(\Lambda _{1}e^{i\theta })\rangle |Coh(\Lambda _{2}e^{-i\theta
})\rangle  \notag \\
&=&\sqrt{1-|\Lambda |^{2}}\sum_{n=0}^{\infty }\Lambda ^{n}\lvert \Delta
_{n}\rangle \lvert \Delta _{n}\rangle .  \label{state_02}
\end{eqnarray}%
We have a more general coefficient $\alpha (\theta )$ by the expansion,
\begin{equation}
\alpha (\theta )=\sum_{l}\alpha _{l}e^{il\theta },
\end{equation}%
and$~\alpha _{l}=\frac{1}{2\pi }\int_{0}^{2\pi }d\theta \alpha (\theta
)e^{-il\theta }$. For this case,
\begin{eqnarray}
\Psi &=&\frac{1}{\sqrt{\mathcal{N}(\Lambda )}}\int_{0}^{2\pi }\frac{d\theta
}{2\pi }~(\sum_{l}\alpha _{l}e^{il\theta })~|Coh(\Lambda _{1}e^{i\theta
})\rangle |Coh(\Lambda _{2}e^{-i\theta })\rangle  \notag \\
&=&\frac{1}{\sqrt{\mathcal{N}(\Lambda )}}~\sum_{l}\sum_{n_{1}=0}^{\infty
}\sum_{n_{2}=0}^{\infty }\left( \int_{0}^{2\pi }\frac{d\theta }{2\pi }\alpha
_{l}e^{i(n_{1}-n_{2}+l)\theta }\right) \Lambda _{1}^{n_{1}}\Lambda
_{2}^{n_{2}}\lvert \Delta _{n_{1}}\rangle \lvert \Delta _{n_{2}}\rangle
\notag \\
&=&\frac{1}{\sqrt{\mathcal{N}(\Lambda )}}\sum_{n=0}^{\infty }\Lambda
^{n}\lvert \Delta _{n}\rangle (\sum_{l}\alpha _{l}\Lambda _{2}^{~l}\lvert
\Delta _{n+l}\rangle ).  \label{state 03}
\end{eqnarray}%
Eq. (\ref{state_02}, \ref{state 03}) is the production of the entangled pair
of giant gravitons over the five-spheres.

The reduced density matrix $\rho _{A}=\mathrm{tr}_{\mathcal{H}_{B}}~|\Psi
\rangle \langle \Psi |$ is
\begin{equation}
\rho _{A}=(1-|\Lambda |^{2})\sum_{n=0}^{\infty }|\Lambda |^{2n}\lvert \Delta
_{n}\rangle \langle \Delta _{n}|~.
\end{equation}%
If we make an identification $e^{-\beta }=|\Lambda |^{2}$,$~\beta =1/T_{%
\mathrm{eff}}$, the thermal mixed YT state is
\begin{equation}
(1-e^{-\beta })\sum_{n=0}^{\infty }e^{-\beta n}\lvert \Delta _{n}\rangle
\langle \Delta _{n}|,~\ \ \ \beta =1/T.  \label{state t 01}
\end{equation}

\subsection{Finite $D$ case}

In Sec 2.1, we have a continuous superposition using an angular distribution
function $\alpha (\theta )$, which is a continuous function. In this
section, we make a discretization of the distribution function in Sec 2.1.
In other words, we use a discrete distribution function in this section. In
order to have a finite $D$ version, we use a large $D$ approximation of the
YT states $\lvert \Delta _{n}\rangle $ by the regularized YT states $\lvert
\tilde{\Delta}_{n}\rangle $ as we will discuss in (\ref{YT_}).

The regularized YT states $\lvert \tilde{\Delta}_{n}\rangle$ is as follows.
In order to have a finite $D$ version of the Hilbert space, we use a large $%
D $ approximation of the YT states by using the large $D$ cat states \cite%
{Lin:2020qao}:
\begin{equation}
\Phi _{D,n}(\Lambda )=\frac{1}{\sqrt{\mathcal{N}_{D,n}}}%
\sum_{m=0}^{D-1}|Coh(\Lambda e^{i\frac{2\pi }{D}m})\rangle e^{-i\frac{2\pi }{%
D}mn},  \label{N state cat}
\end{equation}%
for $n=0,...,D-1$, and $\mathcal{N}_{D,n}=\frac{D^{2}|\Lambda |^{2(n-D\left[
n/D\right] )}}{1-|\Lambda |^{2D}}$ where $\left[ n/D\right] $ denotes the
integer part. We have that $\Vert \Phi _{D,n}(\Lambda )\Vert =1$. It is
pointed out in \cite{Lin:2020qao} that $\Phi _{D,n}(\Lambda )$,$~n=0,...,D-1$%
, corresponds to the $n$-th irreducible representation of the group $Z_{D}$,
and hence they are mutually orthogonal to each other \cite{Lin:2020qao}. The
large $D$ cat states converge to the YT states in the large $D$ limit \cite%
{Lin:2020qao}. These Schrodinger cat states approach the one-row Young
tableau states, with fidelity between them asymptotically reaches 1 at large
$D$. These cat states are defined for $n$ smaller than $D$, and $D$ can be
viewed as a cut-off. However, we can extend the definition of the cat states
for $n$ bigger or equal to $D$ by performing a shift symmetry
identification, $\Phi _{D,n+D}(\Lambda )=\Phi _{D,n}(\Lambda )$.

We define the finite $D$ regularized YT states $\lvert \tilde{\Delta}%
_{n}\rangle $ for large $D$,%
\begin{equation}
\lvert \tilde{\Delta}_{n}\rangle :=\lim_{\Lambda \rightarrow 0,\Lambda \neq
0}\frac{|\Lambda |^{n}}{\Lambda ^{n}}\Phi _{D,n}(\Lambda ).
\end{equation}%
$\frac{|\Lambda |^{n}}{\Lambda ^{n}}=e^{-in\theta _{0}}$ is an overal
constant phase factor. The norm is $\Vert \tilde{\Delta}_{n}\Vert =1$. The
inner product of YT and cat state \cite{Lin:2020qao} is $\langle \Delta
_{n}|\Phi _{D,n}(\Lambda )\rangle =\frac{\Lambda ^{n}}{|\Lambda |^{n}}\sqrt{%
1-|\Lambda |^{2D}},0<|\Lambda |<1$. It is proved in \cite{Lin:2020qao} that
\begin{equation}
\underset{D\rightarrow \infty }{\lim }~\lvert \tilde{\Delta}_{n}\rangle
=\lvert \Delta _{n}\rangle .  \label{YT_}
\end{equation}%
We call $\lvert \tilde{\Delta}_{n}\rangle ~$the regularized version of the
state $\lvert \Delta _{n}\rangle $. We define the regularized multimode
coherent states $|Coh_{D}(w)\rangle $ as follows,%
\begin{equation}
|Coh_{D}(w)\rangle :=\sum_{n=0}^{D-1}w^{n}\lvert \tilde{\Delta}_{n}\rangle ,
\label{expansion_}
\end{equation}%
for any $w\in \mathbb{C}$ and $0<|w|<1$. The expansion of it contains only
finite order terms. We call (\ref{expansion_}) the regularized state or the
finite $D$ version of the state $|Coh(w)\rangle $.

We have the following proposition:

\begin{prop}
\label{prop states} The state $|Coh_{D}(w)\rangle $ converges to $%
|Coh(w)\rangle $ in the infinite D limit, i.e.,
\begin{equation}
\underset{D\rightarrow \infty }{\lim }~|Coh_{D}(w)\rangle =|Coh(w)\rangle .
\end{equation}
\end{prop}

{\textbf{\textit{Proof}}}. We have that $|Coh_{D}(w)\rangle
=\sum_{n=0}^{D-1}w^{n}\lvert \tilde{\Delta}_{n}\rangle $. We also have the
relation $\underset{D\rightarrow \infty }{\lim }\lvert \tilde{\Delta}%
_{n}\rangle =\lvert \Delta _{n}\rangle $. Hence, $\underset{D\rightarrow
\infty }{\lim }~|Coh_{D}(w)\rangle =\sum_{n=0}^{\infty }w^{n}\lvert \Delta
_{n}\rangle =|Coh(w)\rangle .${\hfill $\square $}\newline

Now we use the regularized multimode coherent states to produce the
entangled regularized YT states, as the finite $D$ version of Sec. 2.1. Let $%
\Psi \in \mathcal{H}_{A}\otimes \mathcal{H}_{B}$ and let $\Lambda
_{1}\Lambda _{2}=\Lambda $. We denote $\Lambda _{1}=|\Lambda
_{1}|e^{i\varphi _{1}},\Lambda _{2}=|\Lambda _{2}|e^{i\varphi _{2}}$, and $%
\arg (\Lambda _{2}/\Lambda _{1})=\varphi _{2}-\varphi _{1}$. In the finite $%
D $ version, we have that $\theta _{m}=\frac{2\pi }{D}m$ where $D$ is a
positive integer, and $\lim_{D\rightarrow \infty }\frac{1}{D}%
\sum_{m=0}^{D-1}=\frac{1}{2\pi }\int_{0}^{2\pi }d\theta $.

The finite $D$ version of the entangled multimode coherent states is
\begin{equation}
\Psi =\frac{1}{\sqrt{\mathcal{N}_{D}(\Lambda )}}\frac{1}{D}\sum_{m=0}^{D-1}%
\tilde{\alpha}(m)|Coh_{D}(\Lambda _{1}e^{i\frac{2\pi }{D}m})\rangle
|Coh_{D}(\Lambda _{2}e^{-i\frac{2\pi }{D}m})\rangle ,
\end{equation}%
where $\mathcal{N}_{D}(\Lambda )=\frac{1-|\Lambda |^{2D}}{1-|\Lambda |^{2}}%
.~ $For $\tilde{\alpha}(m)=1,$%
\begin{equation}
\Psi =\frac{1}{\sqrt{\mathcal{N}_{D}(\Lambda )}}\sum_{n=0}^{D-1}\Lambda
^{n}\lvert \tilde{\Delta}_{n}\rangle \lvert \tilde{\Delta}_{n}\rangle .
\label{state_03}
\end{equation}%
This has a natural $D\rightarrow \infty $ limit to (\ref{state_02}). For $%
\tilde{\alpha}(m)=\sum_{l=0}^{D-1}\alpha _{l}e^{il\frac{2\pi }{D}m},$
\begin{equation}
\Psi =\frac{1}{\sqrt{\mathcal{N}_{D}(\Lambda )}}\sum_{n=0}^{D-1}\Lambda
^{n}\lvert \tilde{\Delta}_{n}\rangle (\sum_{l}\alpha _{l}\Lambda
_{2}^{~l}\lvert \tilde{\Delta}_{n+l}\rangle ).  \label{state_04}
\end{equation}

The reduced density matrix $\rho _{A}=\mathrm{tr}_{\mathcal{H}_{B}}~|\Psi
\rangle \langle \Psi |$ is%
\begin{equation}
\rho _{A}=\left( \frac{1-|\Lambda |^{2}}{1-|\Lambda |^{2D}}\right)
\sum_{n=0}^{D-1}|\Lambda |^{2n}\lvert \tilde{\Delta}_{n}\rangle \langle
\tilde{\Delta}_{n}|.~  \label{reduced 03}
\end{equation}

Now we consider the simplification of the states in the $|\Lambda
|\rightarrow 1$ limit. In this limit, by L'Hospital's rule, $\frac{%
(1-|\Lambda |^{2})}{(1-|\Lambda |^{2D})}\rightarrow \frac{1}{D}$, and then (%
\ref{state_03}) is $\Psi =\frac{1}{\sqrt{D}}\sum_{n=0}^{D-1}\lvert \tilde{%
\Delta}_{n}\rangle \lvert \tilde{\Delta}_{n}\rangle $, which is a maximally
entangled state or EPR state \cite{Einstein:1935rr}. For more general $%
|\Lambda |$, the states are non-maximally entangled. The limit $|\Lambda
|\rightarrow 1~$is equivalent to$~|\Lambda _{1}|\rightarrow 1,|\Lambda
_{2}|\rightarrow 1$. Eq. (\ref{state_04}) is then
\begin{equation}
\Psi =\frac{1}{\sqrt{D}}\sum_{n=0}^{D-1}e^{in\varphi _{1}}\lvert \tilde{%
\Delta}_{n}\rangle (\sum_{l}\alpha _{l}e^{i(n+l)\varphi _{2}}\lvert \tilde{%
\Delta}_{n+l}\rangle ),
\end{equation}%
and $|\Psi |=1$, $\sum_{l}|\alpha _{l}|^{2}=1$. We can redefine the basis$%
~e^{in\varphi _{1}}\lvert \tilde{\Delta}_{n}\rangle \rightarrow \lvert
\tilde{\Delta}_{n}\rangle .$ And then,%
\begin{equation}
\Psi =\frac{1}{\sqrt{D}}(I\otimes U)\sum_{n=0}^{D-1}\lvert \tilde{\Delta}%
_{n}\rangle \lvert \tilde{\Delta}_{n}\rangle ,
\end{equation}%
which is in the form of Choi state. $U$ is a unitary matrix of order $D$,%
\begin{equation}
U=\sum_{n^{\prime },n}U_{n^{\prime },n}\lvert \tilde{\Delta}_{n^{\prime
}}\rangle \langle \tilde{\Delta}_{n}|~=\sum_{n^{\prime },n}\sum_{l}\alpha
_{l}e^{i(n+l)(\varphi _{2}-\varphi _{1})}\delta _{n^{\prime },n+l}\lvert
\tilde{\Delta}_{n^{\prime }}\rangle \langle \tilde{\Delta}_{n}|.
\end{equation}%
The norm of $\Psi $ is 1, meaning that $\sum_{n=0}^{D-1}U_{n^{\prime
},n}U_{n,n^{\prime }}^{\ast }=1,~\forall n^{\prime }$, or equivalently $%
UU^{\dagger }=I$. There is a shift symmetry identification $\lvert \tilde{%
\Delta}_{n+D}\rangle \rightarrow \lvert \tilde{\Delta}_{n}\rangle $ for the
states $\lvert \tilde{\Delta}_{n+D}\rangle $ where $n+D>D$.

In the case $\varphi _{2}-\varphi _{1}=0,$ define the matrix
\begin{equation}
X:=\sum_{n^{\prime },n}\delta _{n^{\prime },n+1}\lvert \tilde{\Delta}%
_{n^{\prime }}\rangle \langle \tilde{\Delta}_{n}|,  \label{X_}
\end{equation}%
with the shift symmetry identification $\lvert \tilde{\Delta}_{n+D}\rangle
=\lvert \tilde{\Delta}_{n}\rangle $. $X$ is a cyclic permutation sending $(%
\tilde{\Delta}_{0},\tilde{\Delta}_{1},...,\tilde{\Delta}_{D-2},\tilde{\Delta}%
_{D-1})$ to $(\tilde{\Delta}_{1},\tilde{\Delta}_{2},...,\tilde{\Delta}_{D-1},%
\tilde{\Delta}_{0})$. Then%
\begin{equation}
X_{l}:=X^{l}=\sum_{n^{\prime },n}\delta _{n^{\prime },n+l}\lvert \tilde{%
\Delta}_{n^{\prime }}\rangle \langle \tilde{\Delta}_{n}|,  \label{X_l}
\end{equation}%
$l=0,...,D-1$. $X_{l}$ are also cyclic permutations. The $X_{l}$ are in fact
circulant matrices, and they form a group ring, in which $%
X_{l_{1}}X_{l_{2}}=X_{l_{1}+l_{2}}$. We include $X_{0}=I$ which is the
identity. Hence, $U=\sum_{l}\alpha _{l}X_{l}$. The above $U$ is unitary
since $\sum_{l}|\alpha _{l}|^{2}=1$.

For $\varphi _{2}-\varphi _{1}\neq 0$, the factor $e^{i(n+l)(\varphi
_{2}-\varphi _{1})}=\omega ^{(n+l)}$ corresponds to multiplying a diagonal
unitary matrix $Y=\mathrm{diag}(1,\omega ,\omega ^{2},...,\omega ^{D-1})$,
where $\omega =e^{i(\varphi _{2}-\varphi _{1})}$. Hence,%
\begin{equation}
U=Y\sum_{l}\alpha _{l}X_{l}.
\end{equation}

Define $W=\mathrm{diag}(1,\omega ,\omega ^{2},...,\omega ^{D-1})$, where $%
\omega =e^{i\frac{2\pi }{D}}$. We have that
\begin{equation}
W:=\sum_{n}e^{i\frac{2\pi }{D}n}\lvert \tilde{\Delta}_{n}\rangle \langle
\tilde{\Delta}_{n}|,~~~\ \ \ ~W_{j}:=W^{j}=\sum_{n}e^{i\frac{2\pi }{D}%
nj}\lvert \tilde{\Delta}_{n}\rangle \langle \tilde{\Delta}_{n}|.  \label{W_}
\end{equation}%
We include $W_{0}=I$ which is the identity.

The finite $D$ version of the states is good, in that, it enables to take a $%
|\Lambda |\rightarrow 1$ limit without any divergence that might encounter
for the infinite $D$ case. The finite $D$ case is like a regularization for
the infinite $D$ case.

For the finite $D$ states, we can take two limits of them, one is the
infinite $D$ limit, and another is the $|\Lambda |\rightarrow 1$ limit. This
can be illustrated as follows, with the example of $\frac{1-|\Lambda |^{2}}{%
1-|\Lambda |^{2D}}.$ For finite $D$, $|\Lambda |\rightarrow 1$, $\frac{%
1-|\Lambda |^{2}}{1-|\Lambda |^{2D}}\rightarrow \frac{1}{D}$.~For $%
D\rightarrow \infty $, $|\Lambda |<1$, $\frac{1-|\Lambda |^{2}}{1-|\Lambda
|^{2D}}\rightarrow 1-|\Lambda |^{2}$. In the joint limit, $D\rightarrow
\infty $,$~|\Lambda |\rightarrow 1$,$~\frac{1-|\Lambda |^{2}}{1-|\Lambda
|^{2D}}\rightarrow 0$.

\subsection{More general cases of angular distributions}

In this section, we focus on more general angular distributions, than that
of Sec 2.2. Consider that the distribution function $\alpha (\theta ,\tilde{%
\theta})$ has two angular variables $(\theta ,\tilde{\theta})$ corresponding
to two independent angles of the pair of multi-mode coherent states. The
distribution function $\alpha (\theta ,\tilde{\theta})$ is%
\begin{equation}
\alpha (\theta ,\tilde{\theta})=\sum_{l}\alpha _{l}(\tilde{\theta}%
)e^{il\theta },~~~~~\alpha _{l}(\tilde{\theta})=\frac{1}{2\pi }%
\int_{0}^{2\pi }d\theta \alpha (\theta ,\tilde{\theta})e^{-il\theta }
\end{equation}%
and%
\begin{equation}
\Psi =\frac{1}{\sqrt{\mathcal{N}(\Lambda )}}\int_{0}^{2\pi }\int_{0}^{2\pi }%
\frac{d\theta d\tilde{\theta}}{(2\pi )^{2}}(\sum_{l}\alpha _{l}(\tilde{\theta%
})e^{il\theta })~|Coh(|\Lambda _{1}|e^{i\theta })\rangle |Coh(|\Lambda
_{2}|e^{i(\tilde{\theta}-\theta )})\rangle .
\label{state_general_distribution}
\end{equation}

We also have $\mathrm{\arg }(\Lambda _{2}/\Lambda _{1})=\tilde{\theta}$.
Hence, in the finite $D$ version, $\tilde{\theta}_{j}=\frac{2\pi }{D}j$, and
$\lim_{D\rightarrow \infty }\frac{1}{D}\sum_{j=0}^{D-1}=\frac{1}{2\pi }%
\int_{0}^{2\pi }d\tilde{\theta}$.~Hence,%
\begin{equation}
\alpha _{l,j}=\alpha _{l}(\frac{2\pi }{D}j).
\end{equation}%
We then obtain the finite $D$ version (\ref{state 08}) in the following
proposition \ref{prop states 02}.

\begin{prop}
\label{prop states 02} The finite $D$ version of (\ref%
{state_general_distribution}), i.e.,
\begin{equation}
\Psi =\frac{1}{\sqrt{\mathcal{N}_{D}(\Lambda )}}\frac{1}{D^{2}}%
\sum_{m=0}^{D-1}\sum_{j=0}^{D-1}(\sum_{l}\alpha _{l,j}e^{il\frac{2\pi }{D}%
m})|Coh_{D}(|\Lambda _{1}|e^{i\frac{2\pi }{D}m})\rangle |Coh_{D}(|\Lambda
_{2}|e^{i\frac{2\pi }{D}(j-m)})\rangle  \label{state 08}
\end{equation}%
in the $|\Lambda _{1}\Lambda _{2}|=|\Lambda |\rightarrow 1$ limit, can be
written as%
\begin{equation}
\Psi =\frac{1}{\sqrt{D}}\sum_{n=0}^{D-1}\sum_{l,j}\alpha _{l,j}(I\otimes
U_{l,j})\lvert \tilde{\Delta}_{n}\rangle \lvert \tilde{\Delta}_{n}\rangle ,
\end{equation}%
where $U_{l,j}$ is an unitary matrix, which is $U_{l,j}=W_{j}X_{l}.$
\end{prop}

{\textbf{\textit{Proof}}}.
\begin{equation}
\Psi =\frac{1}{\sqrt{\mathcal{N}_{D}(\Lambda )}}\sum_{n=0}^{D-1}|\Lambda
_{1}|^{n}\lvert \tilde{\Delta}_{n}\rangle (\sum_{j=0}^{D-1}\sum_{l}\alpha
_{l,j}|\Lambda _{2}|^{n+l}e^{i(n+l)\frac{2\pi }{D}j}\lvert \tilde{\Delta}%
_{n+l}\rangle ).
\end{equation}%
We then take $|\Lambda |\rightarrow 1$ limit. Hence define
\begin{eqnarray}
U &=&\sum_{l,j}\alpha _{l,j}W_{j}X_{l}=\sum_{l,j}\alpha _{l,j}U_{l,j}  \notag
\\
&=&\sum_{l,j}\alpha _{l,j}\sum_{n^{\prime },n}e^{i\frac{2\pi }{D}n^{\prime
}j}\delta _{n^{\prime },n+l}\lvert \tilde{\Delta}_{n^{\prime }}\rangle
\langle \tilde{\Delta}_{n}|.~
\end{eqnarray}%
We have that%
\begin{equation}
\Psi =\frac{1}{\sqrt{D}}\sum_{n=0}^{D-1}\sum_{l,j}\alpha _{l,j}(I\otimes
U_{l,j})\lvert \tilde{\Delta}_{n}\rangle \lvert \tilde{\Delta}_{n}\rangle .
\end{equation}%
$U_{l,j}$ form a complete basis for unitary matrices of order $D$. {\hfill $%
\square $}\newline

Due to the orthogonality of the generators $U_{l,j}$, the normalization
condition here is $\sum_{l,j}|\alpha _{l,j}|^{2}=1$. These states are
maximally entangled EPR states \cite{Einstein:1935rr}. They form basis
states for entangled pair of qudits. For more general $|\Lambda |$, the
states are non-maximally entangled. We can also write%
\begin{eqnarray}
\Psi &=&\sum_{l,j}\alpha _{l,j}\Psi _{l,j},~~~~~~~~  \notag \\
\Psi _{l,j} &=&(I\otimes U_{l,j})\Psi _{0},
\end{eqnarray}%
where $\Psi _{0}=\frac{1}{\sqrt{D}}\sum_{n=0}^{D-1}\lvert \tilde{\Delta}%
_{n}\rangle \lvert \tilde{\Delta}_{n}\rangle $.~We include $U_{0,0}=I$ which
is the identity.

The states $\sum_{n}s_{n}\lvert \tilde{\Delta}_{n}\rangle $, with $%
\sum_{n}|s_{n}|^{2}=1$ can be viewed as cat qudits, with qudit
dimensionality $D$. The $D=2$ case and $D=4$ case are particularly
interesting. Since $D=2$ are qubits and can be realized by ordinary cat
qubits, parametrized by amplitudes $\Lambda, -\Lambda $. The $D=4$ can be
realized by superpositions of two different cat qubits, parametrized
together by amplitudes $\Lambda ,\Lambda e^{\frac{i\pi }{2}},-\Lambda
,-\Lambda e^{\frac{i\pi }{2}}.$

We have that $\sum_{l,j}\alpha _{l,j}U_{l,j}$ can be viewed as Choi matrices
for the pair of qudits. For example,%
\begin{equation}
D=2,~X=\sigma _{X},~W=\sigma _{Z},
\end{equation}%
and%
\begin{equation}
D=3,~X=\left[
\begin{array}{ccc}
0 & 0 & 1 \\
1 & 0 & 0 \\
0 & 1 & 0%
\end{array}%
\right] ,~W=\left[
\begin{array}{ccc}
1 & 0 & 0 \\
0 & e^{i\frac{2\pi }{3}} & 0 \\
0 & 0 & e^{i\frac{4\pi }{3}}%
\end{array}%
\right] .
\end{equation}%
Note that the $D=2$ case is Schrodinger cat states with two components, i.e.
ordinary cat qubits. We can perform unitary operations and quantum
operations on these qudits.

On the other hand, the summation (\ref{state 08}) goes over to an
integration (\ref{state_general_distribution}). We can view (\ref%
{state_general_distribution}) as the large $D$ limit of (\ref{state 08}).
The finite-dimension cut-off can be viewed as a regularization for a finite
dimension of the Hilbert space. In the context of gauge/gravity duality, we
can naturally generate entangled pairs of giant graviton qudits and coherent
states, as in Sec. 2. The state $\sum_{n}(I\otimes U)\lvert \Delta
_{n}\rangle \lvert \Delta _{n}\rangle $ is the state of a pair of entangled
fluctuating giant gravitons. The superposition of bipartite entangled
multi-mode coherent states gives rise to Choi states in the Young tableau
basis, with a general unitary $U$. This $U$ can be viewed as a quantum
operation, by the channel/state correspondence \cite{Choi,Jamiolkowski
02,Choi 02}.

After the interaction of the entangled giant graviton states with the
background closed string states, the entangled pair becomes mixed entangled
pair, as we shall discuss in Sec. 5. Now we turn to the relation to Sec.
2.1. Consider the mixed-state entangled state
\begin{equation}
\rho _{AB}=\sum_{l,j}p_{l,j}\lvert \Psi _{l,j}\rangle \langle \Psi _{l,j}|,
\label{state p 02}
\end{equation}%
where $\sum_{l,j}p_{l,j}=1$. For $\rho _{AB}$ as pure states, either
maximally entangled or non-maximally entangled, the entanglement entropy
between the two halves of the five-sphere is $S(\rho _{A})$ or $S(\rho _{B})$%
. For $\rho _{AB}$ as mixed states, the mutual information $I(A,B)$ between
the two halves of the five-sphere is $S(\rho _{A})+S(\rho
_{B})+\sum_{l,j}p_{l,j}\ln p_{l,j}$. We also consider the case when $\rho
_{AB}$ are mixed states, under noisy channels due to interactions with
background fluctuation modes or environment states. These mixed entangled
states can be generated by noisy channels, as we discuss in Sec. 5.

\section{Bipartite entangled states with YT states}

\label{sec 3} \renewcommand{\theequation}{3.\arabic{equation}} %
\setcounter{equation}{0} \renewcommand{\thethm}{3.\arabic{thm}} %
\setcounter{thm}{0} \renewcommand{\theprop}{3.\arabic{prop}} %
\setcounter{prop}{0}

In this section, we consider entangled states which are entangled between
trace states and YT states. We produce entanglement between trace states,
which are dual to closed string states, and YT coherent states. We consider
unitary operations and entangling gates by composite operations of squeezer,
beam splitter, displacer, as well as other operations.

The $\Delta _{n,k}$ is a Young tableau with $n$ columns each with a
column-length $k$. Consider the creation and annihilation operators $%
A_{k}^{\dagger },A_{k}$ on the multi-row Young tableau states $|\Delta
_{n,k}\rangle $,
\begin{equation}
A_{k}^{\dagger }|\Delta _{n,k}\rangle =\sqrt{k(n+1)}|\Delta _{n+1,k}\rangle
,\quad \quad \quad A_{k}|\Delta _{n,k}\rangle =\sqrt{kn}|\Delta
_{n-1,k}\rangle ,  \label{crea. and annih. 02}
\end{equation}%
where $\frac{1}{k}[A_{k},A_{k}^{\dagger }]=1$ and $A_{k}|\Delta
_{0,k}\rangle =0$. The first equation in (\ref{crea. and annih. 02}) can
also be written as%
\begin{equation}
A_{k}^{\dagger }|\Delta _{n-1,k}\rangle =\sqrt{kn}|\Delta _{n,k}\rangle .
\end{equation}%
In other words, $\frac{1}{\sqrt{k}}A_{k}^{\dagger }$ and $\frac{1}{\sqrt{k}}%
A_{k}$ play the role of ordinary creation and annihilation operators, with
the $\frac{1}{\sqrt{k}}$ factor due to our particular convention of the
definition. This is in the large $N$ limit. Hence $|\Delta _{n,k}\rangle =%
\frac{(A_{k}^{\dagger })^{n}}{\sqrt{k^{n}n!}}|\Delta _{0,k}\rangle $ and $%
\Vert \Delta _{n,k}\Vert =1$. The action of $A_{k}^{\dagger }$ is adding one
column of length-$k$ on the Young tableau. The action of $A_{k}$ is removing
one column of length-$k$ from the Young tableau. These creation and
annihilation operators are derived from correlation functions in the large $N
$ quantum field theory, e.g. \cite%
{Berenstein:2017abm,Dhar:2005fg,Koch:2008cm,Lin:2020qao,Balasubramanian:2018yjq}
and references therein.

The coherent states of multi-row ($k$-row) Young tableaux are,%
\begin{equation}
|\Lambda \rangle _{k}=e^{-\frac{|\Lambda |^{2}}{2k}}e^{\frac{1}{k}\Lambda
A_{k}^{\dagger }}|\Delta _{0,k}\rangle =e^{-\frac{|\Lambda |^{2}}{2k}%
}\sum\limits_{n=0}^{\infty }\frac{1}{\sqrt{k^{n}n!}}(\Lambda )^{n}|\Delta
_{n,k}\rangle .  \label{coh. YT 02}
\end{equation}%
We have that $A_{k}|\Lambda \rangle _{k}=\Lambda |\Lambda \rangle _{k}$ and $%
\Vert |\Lambda \rangle _{k}\Vert =1$. Here the range of $\Lambda $ is $%
0<|\Lambda |<\infty $. \ The mean column length is $\langle \hat{N}%
_{k}\rangle _{|\Lambda \rangle _{k}}=\frac{|\Lambda |^{2}}{k}$, and the
variance of the column length is $(\Delta N)_{|\Lambda \rangle _{k}}=\frac{%
|\Lambda |}{\sqrt{k}}$. Hence it is a blob with area $|\Lambda |^{2}/k$, in
the unit of $2\pi \hbar $, saturating the Heisenberg uncertainty relation. $%
|\Lambda |^{2}$ measures the number of boxes of the Young tableau. It is a
blob with $k$ fermions occupied in that droplet in the phase space plane
\cite%
{Lin:2004nb,Berenstein:2004kk,Corley:2001zk,Berenstein:2005aa,Skenderis:2007yb, Balasubramanian:2005mg,Berenstein:2017abm}%
.

For a general $k$, we have the cat state
\begin{equation}
|cat_{\pm }(\Lambda )\rangle _{k}=\frac{1}{\sqrt{N_{k,\pm }}}(|\Lambda
\rangle _{k}\pm |-\Lambda \rangle _{k})  \label{cat 02}
\end{equation}%
where $|cat_{\pm }(\Lambda )\rangle _{k}$ has unit norm and here $N_{k,\pm
}=2(1\pm e^{-\frac{2}{k}|\Lambda |^{2}})$. It is a cat that lives
simultaneously in two different locations on the five-sphere. The
interpretation of the cat state (\ref{cat 02}) is that it is a superposition
of two blobs in the gravity side. This parallels the $D=2$ situation in Sec.
2.3. The two states can be easily distinguished by a parity measurement,
since they have distinct eigenvalues of parity.

The cat states can be approximated by squeezed states after subtraction of
even and odd number of particles:
\begin{eqnarray}
|cat_{+}(\Lambda )\rangle _{k} &\sim &\hat{S}(r)|\Delta _{0,k}\rangle ,
\notag \\
|cat_{-}(\Lambda )\rangle _{k} &\sim &\hat{S}(r)|\Delta _{1,k}\rangle ,
\label{cat approx 02}
\end{eqnarray}%
where in this case the squeeze operator is $\hat{S}(r)=e^{\frac{r}{2k}%
(A_{k}^{\dagger }A_{k}^{\dagger }-A_{k}A_{k})}$. We have generalized them to
new states defined using Young tableaux, from those happen in quantum
information theory \cite{Nielsen etal,Andersen etal} and references therein.
Since $\hat{S}(r)^{\dagger }\hat{S}(r)=I$, the two states (\ref{cat approx
02}) are orthogonal. One can approximate the odd cat as $|cat_{-}(\Lambda
)\rangle \propto A_{k}\hat{S}(r)|\Delta _{0,k}\rangle \propto \hat{S}%
(r)|\Delta _{1,k}\rangle $. There are multiple ways to approximate the even
cat states. One way is by (\ref{cat approx 02}). Another way is to
approximate the even cat as $|cat_{+}(\Lambda )\rangle \propto
A_{k}|cat_{-}(\Lambda )\rangle \propto A_{k}^{~2}\hat{S}(r)|\Delta
_{0,k}\rangle $ and$~A_{k}^{~2}\hat{S}(r)|\Delta _{0,k}\rangle \propto \hat{S%
}(r)|\Delta _{0,k}\rangle +\sqrt{2}\tanh r\hat{S}(r)|\Delta _{2,k}\rangle $.
These approximations of states have high fidelity \cite{Nielsen
etal,Andersen etal}.

We use $\hat{S}(r)^{\dagger }A_{k}\hat{S}(r)=\cosh r~A_{k}+\sinh
r~A_{k}^{\dagger }$ and $\hat{S}(r)^{\dagger }A_{k}^{\dagger }\hat{S}%
(r)=\cosh r~A_{k}^{\dagger }+\sinh r~A_{k}$. This is also a form of
Bogouliubov transformation. We have%
\begin{eqnarray}
\Lambda ^{-1}A_{k}~|cat_{\pm }(\Lambda )\rangle &=&\sqrt{\frac{N_{k,\mp }}{%
N_{k,\pm }}}|cat_{\mp }(\Lambda )\rangle , \\
\hat{P}|cat_{\pm }(\Lambda )\rangle &=&\pm |cat_{\pm }(\Lambda )\rangle ,
\end{eqnarray}%
where $\hat{P}=\exp (i\frac{\pi }{k}A_{k}^{\dagger }A_{k})$. We have that
\begin{eqnarray}
~A_{k}~\hat{S}(r)|\Delta _{0,k}\rangle ~ &=&\sqrt{k}\sinh (r)\hat{S}%
(r)|\Delta _{1,k}\rangle . \\
A_{k}^{\dagger }~\hat{S}(r)|\Delta _{0,k}\rangle ~ &=&\sqrt{k}\cosh (r)\hat{S%
}(r)|\Delta _{1,k}\rangle .
\end{eqnarray}%
For the simplicity of the calculation, we approximate the even cat state as (%
\ref{cat approx 02}). In the approximation for the states, $\sinh (r)\simeq
\sqrt{\frac{N_{k,-}}{N_{k,+}}}\frac{\Lambda }{\sqrt{k}}$. These
approximations simplify calculations and are practically useful.

The beam-splitter operator describes the absorption and emission of closed
strings $|{t_{k}}\rangle $ by giant gravitons with column length $k$. It is
\begin{equation}
\hat{U}(c)=\exp (c(a_{k}^{\dagger }A_{k}-a_{k}A_{k}^{\dagger }))\simeq
1+c(a_{k}^{\dagger }A_{k}-a_{k}A_{k}^{\dagger }).  \label{operation 02}
\end{equation}%
$c$ is proportional to the probability amplitude of the absorption and
emission. The $a_{k}$ is defined in Appendix A for the trace states. The
second equation is when $c$ is small. $a_{k}A_{k}^{\dagger }~$describes the
absorption and $a_{k}^{\dagger }A_{k}$ describes the emission.

We use the operation of $\hat{U}(c)\hat{S}(r)$. The state before the action
of the unitary operations is%
\begin{equation}
|{t_{k}^{0}}\rangle \hat{S}(r)|\Delta _{0,k}\rangle .
\end{equation}%
After the unitary operations, where $c$ is small,%
\begin{eqnarray}
&&\hat{U}(c)(|{t_{k}^{0}}\rangle \hat{S}(r)|\Delta _{0,k}\rangle )  \notag \\
&\simeq &\frac{1}{\sqrt{\mathcal{N}}}\left( |{t_{k}^{0}}\rangle \hat{S}%
(r)|\Delta _{0,k}\rangle +c\sqrt{k}\sinh (r)|{t_{k}^{1}}\rangle \hat{S}%
(r)|\Delta _{1,k}\rangle \right) \\
&\simeq &\frac{1}{\sqrt{\mathcal{N}}}\left( |{t_{k}^{0}}\rangle
|cat_{+}(\Lambda )\rangle +ck\sinh (r)\frac{|{t_{k}^{1}}\rangle }{\sqrt{k}}%
|cat_{-}(\Lambda )\rangle \right) .
\end{eqnarray}%
Here $\frac{1}{\sqrt{\mathcal{N}}}$ is a normalization factor. The two terms
in the superposition are orthogonal to each other. Hence we produce
entangled state%
\begin{equation}
c_{1}|{t_{k}^{0}}\rangle |cat_{+}(\Lambda )\rangle +c_{2}\frac{|{t_{k}^{1}}%
\rangle }{\sqrt{k}}|cat_{-}(\Lambda )\rangle .
\end{equation}%
We can tune the parameters $ck\sinh (r)$ to obtain arbitrary $\frac{c_{2}}{%
c_{1}}=ck\sinh (r)$.

There are also other methods of producing the entangled states, such as by
the operation of $\hat{S}(r)\hat{U}(c)$. We have another method:%
\begin{eqnarray}
\hat{S}(r)\hat{U}(c)(\frac{1}{\sqrt{k}}|{t_{k}^{1}}\rangle |\Delta
_{0,k}\rangle ) &\simeq &\frac{1}{\sqrt{\mathcal{N}}}\left( \frac{1}{\sqrt{k}%
}|{t_{k}^{1}}\rangle \hat{S}(r)|\Delta _{0,k}\rangle -ck|{t_{k}^{0}}\rangle
\hat{S}(r)|\Delta _{1,k}\rangle \right)  \notag \\
&\simeq &\frac{1}{\sqrt{\mathcal{N}}}\left( \frac{|{t_{k}^{1}}\rangle }{%
\sqrt{k}}|cat_{+}(\Lambda )\rangle -ck|{t_{k}^{0}}\rangle |cat_{-}(\Lambda
)\rangle \right) .
\end{eqnarray}%
Similarly, we can produce the states for general $l\geqslant 1$, where $%
\frac{|{t_{k}^{l}}\rangle }{\sqrt{l!k^{l}}}$ has a unit norm. Denote $|\Psi
_{0}\rangle =\frac{|{t_{k}^{l_{1}}}\rangle }{\sqrt{l_{1}!k^{l_{1}}}}|\Delta
_{0,k}\rangle $. We can produce the entangled states%
\begin{eqnarray}
&&\hat{S}(r)\hat{U}(c)|\Psi _{0}\rangle  \notag \\
&\simeq &\frac{1}{\sqrt{\mathcal{N}}}\left( \frac{|{t_{k}^{l_{1}}}\rangle }{%
\sqrt{l_{1}!k^{l_{1}}}}|cat_{+}(\Lambda )\rangle -ck\frac{|{t_{k}^{l_{2}}}%
\rangle }{\sqrt{l_{2}!k^{l_{2}}}}|cat_{-}(\Lambda )\rangle \right) ,
\end{eqnarray}%
where $l_{2}=l_{1}-1$. Hence we can produce the entangled states of the form
\begin{equation}
c_{1}|{\phi }_{+}\rangle |cat_{+}(\Lambda )\rangle +c_{2}|{\phi }_{-}\rangle
|cat_{-}(\Lambda )\rangle .  \label{micro macro 02}
\end{equation}%
The two terms have opposite parity.

Starting from the $|cat_{-}(\Lambda )\rangle $ state, by acting the single
subtraction operator $\Lambda ^{-1}A_{k}$ on it, we could produce the $%
|cat_{+}(\Lambda )\rangle $ state. The big coherent state amplitude limit is
particularly interesting, since for $|\Lambda |\gg 1,$ $\frac{N_{k,\mp }}{%
N_{k,\pm }}\simeq 1$, and the calculations can be simplified. We have the
Pauli operators $X,Y,Z$ acting on the cat qubits, $X:=\Lambda ^{-1}A_{k}$,$%
~Y:=i\Lambda ^{-1}A_{k}\exp (i\frac{\pi }{k}A_{k}^{\dagger }A_{k})$, and $%
Z:=\exp (i\frac{\pi }{k}A_{k}^{\dagger }A_{k})$. Here, $Z$ coincides with
the parity operator, since $\hat{P}|cat_{\pm }(\Lambda )\rangle =\pm
|cat_{\pm }(\Lambda )\rangle $. $X$ and $Z$ are bit-flip and phase-flip
operators acting on the cat qubits. They, together with the identity,
generates the Pauli group.

The displacer is $\hat{D}(\alpha )=\exp (\frac{1}{k}(\alpha A_{k}^{\dagger }-%
\bar{\alpha}A_{k}))$. We have $\hat{D}(\alpha )^{\dagger }A_{k}\hat{D}%
(\alpha )=A_{k}+\alpha $ and$~\hat{D}(\alpha )^{\dagger }A_{k}^{\dagger }%
\hat{D}(\alpha )=A_{k}^{\dagger }+\bar{\alpha}$. Moreover, there is a qubit
phase rotation gate. We generalize slightly the gate in \cite{Marek etal}.
We use displaced-two-photon-subtractor
\begin{equation}
\hat{R}=\frac{A_{k}}{\Lambda }\hat{D}(\alpha )^{\dagger }\frac{A_{k}}{%
\Lambda }\hat{D}(\alpha )
\end{equation}%
instead, and%
\begin{equation}
\hat{R}~(c_{1}|cat_{+}(\Lambda )\rangle +c_{2}|cat_{-}(\Lambda )\rangle
)=c_{1}^{\prime }|cat_{+}(\Lambda )\rangle +c_{2}^{\prime }|cat_{-}(\Lambda
)\rangle ,
\end{equation}%
where%
\begin{equation}
\frac{c_{2}^{\prime }}{c_{1}^{\prime }}=\frac{c_{2}}{c_{1}}\frac{1+\frac{%
c_{1}}{c_{2}}\alpha /\Lambda }{1+\frac{c_{2}}{c_{1}}\alpha /\Lambda }.
\end{equation}%
These operations realize phase rotations of the cat qubits.

We can also produce entanglement between trace states and YT coherent
states. We can use the operation $\hat{D}(\Lambda )\hat{U}(c)$. We have$%
~A_{k}^{\dagger }\hat{D}(\Lambda )|\Delta _{0,k}\rangle =\bar{\Lambda}\hat{D}%
(\Lambda )|\Delta _{0,k}\rangle +\hat{D}(\Lambda )|\Delta _{1,k}\rangle $.
Denote $|\Psi _{0}\rangle =\frac{|{t_{k}^{l_{1}}}\rangle }{\sqrt{%
l_{1}!k^{l_{1}}}}|\Delta _{0,k}\rangle $. We can also produce the entangled
states:%
\begin{eqnarray}
&&\hat{D}(\Lambda )\hat{U}(c)|\Psi _{0}\rangle  \notag \\
&\simeq &\frac{1}{\sqrt{\mathcal{N}}}\left( \frac{|{t_{k}^{l_{1}}}\rangle }{%
\sqrt{l_{1}!k^{l_{1}}}}\hat{D}(\Lambda )|\Delta _{0,k}\rangle -ck\frac{|{%
t_{k}^{l_{2}}}\rangle }{\sqrt{l_{2}!k^{l_{2}}}}\hat{D}(\Lambda )|\Delta
_{1,k}\rangle \right) ,  \label{micro macro 03}
\end{eqnarray}%
where $l_{1}-l_{2}=1$. Since $\hat{D}(\Lambda )^{\dagger }\hat{D}(\Lambda
)=I $, $\hat{D}(\Lambda )|\Delta _{0,k}\rangle $ and $\hat{D}(\Lambda
)|\Delta _{1,k}\rangle $ are orthogonal. This type of micro-macro entangled
states can be experimentally realized using photons and coherent states of
photons, e.g. \cite{Lvovsky etal,Bruno etal}. Here, these can be interpreted
as the entangled states of closed string states and the droplets.

Both (\ref{micro macro 02}) and (\ref{micro macro 03}) are methods creating
micro-macro entangled states. These types of cat states and other macro
states as well as their entangled states, have been realized experimentally
in the context of photonic coherent states. The photon subtraction
techniques are crucial in quantum state preparations, e.g. \cite{Nielsen
etal,Andersen etal,Morin etal}.

\section{Mixed YT states and noisy YT states}

\label{sec 4} \renewcommand{\theequation}{4.\arabic{equation}} %
\setcounter{equation}{0} \renewcommand{\thethm}{4.\arabic{thm}} %
\setcounter{thm}{0} \renewcommand{\theprop}{4.\arabic{prop}} %
\setcounter{prop}{0}

\subsection{Mixed states of YT}

Now in this section, we consider mixed states of giant gravitons arising
from ensemble of pure states. In this section, we consider mixed states of
multimode coherent states and YT basis. The mixed states of multi-mode
coherent states is
\begin{equation}
\rho =\rho \lbrack p(\theta )]:=\frac{1}{\mathcal{N}_{D}(\Lambda )}%
\int_{0}^{2\pi }d\theta ~p(\theta )~|Coh(\Lambda e^{i\theta })\rangle
\langle Coh(\Lambda e^{i\theta })|,  \label{mixed state 02}
\end{equation}%
where $p(\theta )$ is a distribution function, in the angular direction. We
have $p(\theta )\geq 0$ and $\int_{0}^{2\pi }d\theta ~p(\theta )=1$, and
hence $\mathrm{tr}~\rho =1$. $\rho \lbrack p(\theta )]$ is a functional of $%
p(\theta )$. The coefficient $(1-|\Lambda |^{2})$ is a normalization factor
for the un-normalized multi-mode coherent states (\ref{Coh 01}). We will
expand it in the Young tableau basis. We will show that angular
superpositions of multi-mode coherent state density matrices, give rise to
mixed states in the Young tableau basis.

In \cite{Lin:2020qao} superposition pure states of the multi-mode coherent
states were analyzed, while here we analyze mixed states of the ensemble of
the multi-mode coherent states, by superimposing density matrices.

For $p(\theta )=\frac{1}{2\pi },$%
\begin{eqnarray}
\rho _{0} &=&(1-|\Lambda |^{2})\frac{1}{2\pi }\int_{0}^{2\pi }d\theta
\sum_{n_{1}=0}^{\infty }\sum_{n_{2}=0}^{\infty }e^{i(n_{1}-n_{2})\theta
}|\Lambda |^{n_{1}+n_{2}}\lvert \Delta _{n_{1}}\rangle \langle \Delta
_{n_{2}}|  \notag \\
&=&(1-|\Lambda |^{2})\sum_{n=0}^{\infty }|\Lambda |^{2n}\lvert \Delta
_{n}\rangle \langle \Delta _{n}|.  \label{infinite D_01}
\end{eqnarray}%
It is a mixed state with `thermal'-like distribution. It can be viewed as a
Gibbs state. If we make an identification $e^{-\beta }=|\Lambda |^{2}$,$%
~\beta =1/T_{\mathrm{eff}}$, the thermal mixed YT state is%
\begin{equation}
(1-e^{-\beta })\sum_{n=0}^{\infty }e^{-\beta n}\lvert \Delta _{n}\rangle
\langle \Delta _{n}|,~\ \ \ \beta =1/T.  \label{state_07}
\end{equation}

More generally, we have $p(\theta )=c_{0}\frac{1}{2\pi }+\sum_{l\geq 1}c_{l}%
\frac{1}{\pi }\cos ^{2}\frac{1}{2}l\theta =\frac{1}{2\pi }+\sum_{l\geq 1}%
\frac{c_{l}}{2\pi }\cos l\theta $, where $\sum_{l\geq 0}c_{l}=1$ and$%
~c_{l}\geq 0,\forall l\in \mathbb{Z}_{>0}$. The above distribution functions
are all linear combinations of periodic functions, with periods $\frac{2\pi
}{l}$, along the circle. This is a deformed distribution with respect to the
constant distribution $p(\theta )=\frac{1}{2\pi }$. From (\ref{mixed state
02}) we have that
\begin{equation}
\rho =(1-|\Lambda |^{2})\sum_{n=0}^{\infty }|\Lambda |^{2n}(\lvert \Delta
_{n}\rangle \langle \Delta _{n}|+\sum_{l\geq 1}\frac{c_{l}}{2}(\lvert \Delta
_{n+l}\rangle \langle \Delta _{n}|+\lvert \Delta _{n}\rangle \langle \Delta
_{n+l}|)).  \label{infinite D_02}
\end{equation}%
The second piece are off-diagonal components of the density matrix and $%
\mathrm{tr}~\rho =1$. The number of non-zero $c_{l}$ can be finite.

These mixed states are smeared distribution of giant gravitons over the
five-sphere. Mixed states of giant gravitons and/or droplet configurations,
and their related aspects, were also considered in e.g. \cite%
{Balasubramanian:2005mg},\cite{Simon:2018laf},\cite{Lin:2020qao},\cite%
{Berenstein:2018lrm}--\cite{Balasubramanian:2007zt}.

\subsection{Finite $D$ case}

In Sec 4.1, we have a continuous distribution function $p(\theta )$ to
generate the mixed states. In this section, we make a discretization of the
distribution function in Sec 4.1. In other words, we use a discrete
distribution function in this section. We have that $\theta _{m}=\frac{2\pi
}{D}m$ where $D$ is a positive integer, and $\lim_{D\rightarrow \infty }%
\frac{1}{D}\sum_{m=0}^{D-1}=\frac{1}{2\pi }\int_{0}^{2\pi }d\theta $. The
density matrix of the mixed state is
\begin{equation}
\rho =\rho \lbrack \{p_{m}\}]:=\frac{1}{\mathcal{N}_{D}(\Lambda )}\frac{2\pi
}{D}\sum_{m=0}^{D-1}~p_{m}|Coh_{D}(\Lambda e^{i\frac{2\pi }{D}m})\rangle
\langle Coh_{D}(\Lambda e^{i\frac{2\pi }{D}m})|,
\end{equation}%
where $\mathcal{N}_{D}(\Lambda )=\frac{(1-|\Lambda |^{2})}{(1-|\Lambda
|^{2D})}$. We have that $\frac{2\pi }{D}\sum_{m=0}^{D-1}p_{m}=1$.

For $p_{m}=\frac{1}{2\pi },$%
\begin{equation}
\rho _{0}=\frac{(1-|\Lambda |^{2})}{(1-|\Lambda |^{2D})}\sum_{n=0}^{D-1}|%
\Lambda |^{2n}\lvert \tilde{\Delta}_{n}\rangle \langle \tilde{\Delta}_{n}|
\label{density_05}
\end{equation}%
and $\mathrm{tr~}\rho _{0}=1$. This is the finite $D$ version of (\ref%
{infinite D_01}), and has a natural $D\rightarrow \infty $ limit to (\ref%
{infinite D_01}). Eq. (\ref{density_05}) can also be obtained by partial
tracing as in (\ref{reduced 03}).

For $p_{m}=c_{0}\frac{1}{2\pi }+\sum_{l=1}^{[\frac{D}{2}]}c_{l}\frac{1}{\pi }%
\cos ^{2}l\frac{\pi }{D}m$, where $\sum_{l\geq 0}c_{l}=1$ and $c_{l}\geq
0,\forall l$,%
\begin{eqnarray}
\rho &=&\rho _{0}+\frac{1}{\mathcal{N}_{D}(\Lambda )D}(\sum_{l=1}^{[\frac{D}{%
2}]}\sum_{m=0}^{D-1}\sum_{n_{1},n_{2}}\frac{c_{l}}{2}e^{i(n_{1}-n_{2}-l)%
\frac{2\pi }{D}m}|\Lambda |^{n_{1}+n_{2}}\lvert \Delta _{n_{1}}\rangle
\langle \Delta _{n_{2}}|+\mathrm{h.c.})  \notag \\
&=&\rho _{0}+\frac{(1-|\Lambda |^{2})}{(1-|\Lambda |^{2D})}\sum_{l=1}^{[%
\frac{D}{2}]}\frac{c_{l}}{2}\sum_{n=0}^{D-1}|\Lambda |^{2n}(\lvert \tilde{%
\Delta}_{n+l}\rangle \langle \tilde{\Delta}_{n}|+\lvert \tilde{\Delta}%
_{n}\rangle \langle \tilde{\Delta}_{n+l}|)  \notag \\
&=&\rho _{0}+\sum_{l=1}^{[\frac{D}{2}]}\frac{c_{l}}{2}(X_{l}\rho _{0}+\rho
_{0}X_{l}^{\dagger }).  \label{density_06}
\end{eqnarray}%
We have set $c_{l}=0$ for $l>[\frac{D}{2}]$. This is the finite $D$ version
of (\ref{infinite D_02}). In the limit $|\Lambda |\rightarrow 1$, $\rho _{0}=%
\frac{I}{D}$ and $\rho =\frac{1}{D}(I+\sum_{l=1}^{[\frac{D}{2}]}\frac{c_{l}}{%
2}(X_{l}+X_{l}^{\dagger }))$. $X_{l}$ is (\ref{X_l}) and they are traceless
for $1\leqslant l<D$.

The second Renyi entropy $S^{(2)}(\rho )=-\log ($\textrm{tr~}$\rho ^{2})$
and the entropy $S(\rho )=-\mathrm{tr}(\rho \log \rho )$ of (\ref{density_05}%
) and (\ref{density_06}) are:%
\begin{equation}
S^{(2)}(\rho )=-\log \left( \frac{(1-|\Lambda |^{2})(1+|\Lambda |^{2D})}{%
(1-|\Lambda |^{2D})(1+|\Lambda |^{2})}(1+\sum_{1\leqslant l\leqslant \lbrack
\frac{D-1}{2}]}\frac{c_{l}^{2}}{2}+~\frac{(1+|\Lambda |^{D})^{2}}{%
(1+|\Lambda |^{2D})}\sum_{[\frac{D-1}{2}]<l\leqslant \lbrack \frac{D}{2}]}%
\frac{c_{l}^{2}}{2}~)\right)  \label{entropy 05}
\end{equation}%
and%
\begin{equation}
S(\rho )=-\mathrm{tr}(\rho _{0}\log \rho _{0})-2D(\sum_{1\leqslant
l\leqslant \lbrack \frac{D-1}{2}]}\frac{c_{l}}{2D}\log \frac{c_{l}}{2D}%
+~2\sum_{[\frac{D-1}{2}]<l\leqslant \lbrack \frac{D}{2}]}\frac{c_{l}}{2D}%
\log \frac{c_{l}}{2D}),  \label{entropy 06}
\end{equation}%
where $-\mathrm{tr}(\rho _{0}\log \rho _{0})=\log D$,~for the$~|\Lambda
|\rightarrow 1$ case. For odd $D$, the last term in (\ref{entropy 05}) and (%
\ref{entropy 06}) is absent. The second term in (\ref{entropy 06}) also
describes the entropy increase $\Delta S=S(\rho )-S(\rho _{0})$. With the
finite $D$ results, we can take an infinite $D$ limit to obtain the
expressions for Sec 4.1.

We make an identification $e^{-\beta }=|\Lambda |^{2}$,$~\beta =1/T_{\mathrm{%
eff}}$, the thermal mixed YT state with finite $D$ is
\begin{equation}
\frac{(1-e^{-\beta })}{(1-e^{-\beta D})}\sum_{n=0}^{D-1}e^{-\beta n}\lvert
\tilde{\Delta}_{n}\rangle \langle \tilde{\Delta}_{n}|,~\ \ \ \beta =1/T.
\end{equation}%
This is a Gibbs state with a finite dimension. Its infinite $D$ limit goes
over to (\ref{state_07}).

\subsection{Noisy YT channels and noisy YT states}

The mixed noisy YT states can be produced by interactions with background
fluctuation modes which we view as environment states. Now in this section,
we consider producing these mixed states by noisy quantum channels. The
finite dimensional case can be considered as an approximation of the
infinite dimensional case. For finite dimensional cases, we would use the
regularized YT states $\lvert \tilde{\Delta}_{n}\rangle $ to approximate the
YT states $\lvert \Delta _{n}\rangle $. The infinite $D$ limit of the former
gives the later. We denote $A$ the system and $B$ the environment, in this
section.

The state $\lvert \Psi _{AB}^{0}\rangle =\sum_{n}s_{n}\lvert \Delta
_{n}\rangle \lvert \phi _{0}\rangle $, where $\sum_{n}|s_{n}|^{2}=1$,
evolves to%
\begin{eqnarray}
\lvert \Psi _{AB}\rangle &=&U^{AB}\sum_{n}s_{n}\lvert \Delta _{n}\rangle
\lvert \phi _{0}\rangle  \notag \\
&=&\sum_{n,n^{\prime },i}s_{n}c_{n^{\prime },n}^{i}\lvert \Delta _{n^{\prime
}}\rangle \lvert \phi _{i}\rangle ,  \label{evolution_}
\end{eqnarray}%
in which $\lvert \phi _{i}\rangle $ are environment states of $B$ and%
\begin{equation}
c_{n^{\prime },n}^{i,j}=\langle \phi _{i}|\langle \Delta _{n^{\prime
}}|U^{AB}\lvert \Delta _{n}\rangle \lvert \phi _{j}\rangle ,~\ \ \
~~~~c_{n^{\prime },n}^{i}=c_{n^{\prime },n}^{i,0},  \label{interaction_}
\end{equation}%
where $U^{AB}\in U(\mathcal{H}_{AB})$.

Here, $c_{n^{\prime },n}^{i}$ encodes the information of the interaction
between the states and the environment. (\ref{interaction_}) is a four-point
function, and in the special case when $\lvert \phi _{0}\rangle $ is the
empty vacuum state, is a three-point function. For example, we view $\lvert
\Delta _{n}\rangle $ as giant graviton states and $\lvert \phi _{i}\rangle $
as close string states. We interpret the coefficients as four-point
functions. The correlation functions from the field theory side describing
closed strings interacting with giant gravitons were considered in e.g. \cite%
{Bissi:2011dc}--\cite{Vescovi:2021fjf} and references therein.

Such interactions and correlators can be described by the absorbtion and
emission of closed strings by giant gravitons, e.g. \cite{Chen:2019gsb}--%
\cite{Berenstein:2020grg}, \cite{Lin:2012ey} and their related discussions,
in the context of string theory and gauge-string duality. These coefficients
$c_{n^{\prime },n}^{i}$ are determined by the Hamiltonian of the dual CFT or
dual gauge theory. The unitary matrix depends on the details of the
Hamiltonian of the dual field theory.

Eq. (\ref{evolution_}) are entangled states. Note that $\sum_{n}s_{n}\lvert
\Delta _{n}\rangle $ is a more general states like the macro state in (\ref%
{micro macro 02}) in Sec. 3, and $\lvert \phi _{i}\rangle $ is a more
general background closed string states like the micro state in (\ref{micro
macro 02}). $U$ is a more general evolution operator, like the one in (\ref%
{operation 02}). The $U$ in Sec. 3 is a special case. More generally,
\begin{equation}
U^{AB}=\sum_{i,j,n^{\prime },n}c_{n^{\prime },n}^{i,j}\lvert \Delta
_{n^{\prime }}\rangle \lvert \phi _{i}\rangle \langle \phi _{j}|\langle
\Delta _{n}|.
\end{equation}%
Since $\sum_{n^{\prime },i}c_{n,n^{\prime }}^{j,i~\ast }c_{n^{\prime
},l}^{i,k}=\delta _{n,l}\delta _{j,k}$, hence for the $j=0,k=0$ component, $%
\sum_{n^{\prime },i}c_{n,n^{\prime }}^{i~\ast }c_{n^{\prime },l}^{i}=\delta
_{n,l}$.

The system of giant gravitons can be viewed as an open quantum system, since
they can interact with background closed string states. $AB$ is a closed
subsystem of the Hilbert space of the dual CFT. States in $A$ can interact
with states in $B$, hence $A$ can be viewed as an open quantum system, and $%
B $ viewed as environment \cite{Lindblad:1975ef}. The evolution in the close
system $AB$ is unitary \cite{Lindblad:1975ef}. The dimension of the
environment states is $d_{B}$. Taking trace over $B$ is the same as summing
over the basis of environment states $\lvert \phi _{i}\rangle $ in $\mathcal{%
H}_{B}.$

The noisy YT state can be viewed as the transformation of a pure YT state by
a noisy quantum channel defined by a linear completely-positive
trace-preserving map. The noisy YT channel is $\mathcal{E}$,
\begin{equation}
\rho _{A}=\mathcal{E}(\rho _{A}^{0})=\text{\textrm{tr}}_{B}~U^{AB}(\rho
_{A}^{0}\otimes \rho _{B}^{0})U^{AB}{}^{\dagger }.  \label{channel_}
\end{equation}%
In the case that the input are pure states, we have that $\rho
_{A}^{0}=\sum_{n,l}s_{n}\bar{s}_{l}\lvert \Delta _{n}\rangle \langle \Delta
_{l}|$ and $\rho _{B}^{0}=\lvert \phi _{0}\rangle \langle \phi _{0}|.$ The
output state is%
\begin{equation}
\rho _{A}=\sum_{n^{\prime },l^{\prime }}\sum_{n,l,i}s_{n}\bar{s}%
_{l}c_{n^{\prime },n}^{i}c_{l,l^{\prime }}^{i~\ast }\lvert \Delta
_{n^{\prime }}\rangle \langle \Delta _{l^{\prime }}|.
\end{equation}%
It can be written as $\rho _{A}=\sum_{i}E_{i}\rho _{A}^{0}E_{i}^{\dagger }$%
,~where the Kraus operator in this case is $E_{i}=\sum_{n^{\prime
},n}c_{n^{\prime },n}^{i,0}\lvert \Delta _{n^{\prime }}\rangle \langle
\Delta _{n}|$, or $(E_{i})_{n^{\prime },n}=c_{n^{\prime },n}^{i,0}$. On the
other hand, $\rho _{B}=~\mathrm{tr}_{A}~U^{AB}(\rho _{A}^{0}\otimes \rho
_{B}^{0})U^{AB}{}^{\dagger }$ and%
\begin{equation}
\rho _{B}=\sum_{i,j}\sum_{n,l,n^{\prime }}s_{n}\bar{s}_{l}c_{n^{\prime
},n}^{i}c_{l,n^{\prime }}^{j~\ast }\lvert \phi _{i}\rangle \langle \phi
_{j}|.
\end{equation}%
It can also be written as $\rho _{B}=\sum_{i,j}M_{ij}\lvert \phi _{i}\rangle
\langle \phi _{j}|$,$~$where the matrix elements of $M$ is $M_{ij}=~\mathrm{%
tr}(\rho _{A}^{0}E^{j\dagger }E^{i})$. We have also%
\begin{equation}
\rho _{AB}=\sum_{i,j}(E^{i}\rho _{A}^{0}E^{j\dagger })\lvert \phi
_{i}\rangle \langle \phi _{j}|.
\end{equation}%
In this case, $\rho _{B}$ may be understood as being through a conjugate
channel \cite{Holevo,King etal,Devetak etal}.

The increase of the entropy of $A$ is%
\begin{equation}
\Delta S_{A}=S(\rho _{A})-S(\rho _{A}^{0})=-\text{\textrm{tr}}(M\log M).
\end{equation}%
The entropy production, in other words the increase of the entropy of $A$,
after the interaction is $\Delta S_{A}$. This is the entropy production of
the system, after the interaction of the system with the environment. This
is also proportional to the change of the mutual information between $A$ and
$B$, after the interaction. We have that
\begin{equation}
I(A,B)-I(A,B)^{0}=S(\rho _{AB}||\rho _{A}\otimes \rho _{B})=-2~\mathrm{tr}%
(M\log M).
\end{equation}%
We can have these interaction process occur multiple times, and the
evolution process forms a semigroup evolution along the time direction.

The above derivation assumes that $\rho _{A}^{0}$ is pure. The definition of
the channel (\ref{channel_}) is good for both the cases when $\rho _{A}^{0}$
is pure or mixed, since (\ref{channel_}) is a linear map of the input
density matrix. If we trace out the environment states, we hence define a
noisy quantum channel.

Some examples of noisy YT states are as follows. First, the pure YT state is%
\begin{equation}
\lvert \Delta _{n}\rangle \langle \Delta _{n}|.
\end{equation}%
The noisy YT state is
\begin{equation}
\alpha \lvert \Delta _{n}\rangle \langle \Delta _{n}|+\sum_{n^{\prime }\neq
n}p_{n^{\prime }}\lvert \Delta _{n^{\prime }}\rangle \langle \Delta
_{n^{\prime }}|,~~~~\sum_{n^{\prime }\neq n}p_{n^{\prime }}=1-\alpha
,~0<\alpha <1.
\end{equation}%
The thermal mixed YT state is
\begin{equation}
\rho _{0}=(1-e^{-\beta })\sum_{n}e^{-\beta n}\lvert \Delta _{n}\rangle
\langle \Delta _{n}|,~\ \ \ \beta =1/T,  \label{thermal 04}
\end{equation}%
which is also (\ref{state_07}) in Sec. 4.1. The noisy YT state with a
thermal noise is
\begin{equation}
\alpha \lvert \Delta _{n}\rangle \langle \Delta _{n}|+(1-\alpha )\rho
_{0},~~~~0<\alpha <1,  \label{state 09}
\end{equation}%
where $\rho _{0}$ is (\ref{thermal 04}) and the second term in (\ref{state
09}) represents the thermal noise. Some of these noisy states include those
mixed states discussed in Sec. 4.1 and 4.2.

\section{Noisy entangled YT states}

\label{sec 5} \renewcommand{\theequation}{5.\arabic{equation}} %
\setcounter{equation}{0} \renewcommand{\thethm}{5.\arabic{thm}} %
\setcounter{thm}{0} \renewcommand{\theprop}{5.\arabic{prop}} %
\setcounter{prop}{0}

In this section, we produce mixed entangled pair by a noisy channel. The
pure entangled pair becomes mixed entangled pair after interaction with
background environment states. We generalize the situation of Sec 4.3, from
single partite states interacting with the environment states to bipartite
entangled states interacting with the environment states. We denote $AB$ the
entangled pair and $C$ the environment, in this section.

Consider the states having interactions with background fluctuation modes,
which we model as noise. The evolution of the total system takes into
account the interaction hamiltonian and the unitary evolution operators. The
background fluctuation modes are environment states, for example a
background of random closed strings.

The initial state in $AB$ and in $C$ is$~\lvert \Psi _{AB}^{0}\rangle
=\sum_{n}\sqrt{p_{n}}\lvert \Delta _{n}\rangle \lvert \Delta _{n}\rangle $
and $\lvert \Psi _{C}^{0}\rangle =\lvert \phi _{0}\rangle $, where $%
\sum_{n}p_{n}=1$. The initial state of the entangled pair ($AB$) and the
environment ($C$) is
\begin{equation}
\lvert \Psi _{ABC}^{0}\rangle =\sum_{n}\sqrt{p_{n}}\lvert \Delta _{n}\rangle
\lvert \Delta _{n}\rangle \lvert \phi _{0}\rangle .
\end{equation}%
We consider interaction between $B$ and $C$. Under the interaction, the
state $\lvert \Psi _{ABC}^{0}\rangle $ evolves to $\lvert \Psi _{ABC}\rangle
$. We denote the initial states by a subscript `0', such as $\rho
_{AB}^{0},\rho _{C}^{0},\rho _{ABC}^{0}$, and after the evolution, the
states are $\rho _{AB},\rho _{C},\rho _{ABC}$. The initial state is a direct
product state between $AB$ and $C$, but not a direct product state between $%
A $ and $B$, i.e. $\rho _{ABC}^{0}=\rho _{AB}^{0}\otimes \rho _{C}^{0}.$ We
have that $\rho _{A}^{0}=\sum_{n}p_{n}\lvert \Delta _{n}\rangle \langle
\Delta _{n}|=\rho _{B}^{0}$ are mixed states. After the interaction between $%
AB$ and $C$, the state after evolution $\rho _{ABC}$ is not a direct product
state of $AB$ and $C$.

Due to the interaction with the environment,
\begin{eqnarray}
\lvert \Psi _{ABC}\rangle &=&(I\otimes U^{BC})\sum_{n}\sqrt{p_{n}}\lvert
\Delta _{n}\rangle \lvert \Delta _{n}\rangle \lvert \phi _{0}\rangle  \notag
\\
&=&\sum_{n,n^{\prime },i}\sqrt{p_{n}}A_{n^{\prime },n}^{i}\lvert \Delta
_{n}\rangle \lvert \Delta _{n^{\prime }}\rangle \lvert \phi _{i}\rangle ,
\end{eqnarray}%
in which $\lvert \phi _{i}\rangle $ is the orthonormal basis of the
environment states and $U^{BC}\in U(\mathcal{H}_{BC})$. We have that$~\lvert
\Psi _{ABC}\rangle $ has a unit norm.$~$Here,%
\begin{equation}
A_{n^{\prime },n}^{i}=\langle \phi _{i}|\langle \Delta _{n^{\prime
}}|U^{BC}\lvert \Delta _{n}\rangle \lvert \phi _{0}\rangle .
\label{interaction 02}
\end{equation}%
This is equivalent to scattering matrix elements. $p_{n}$ encodes the
information of the bipartite states and $A_{n^{\prime },n}^{i}$ encodes the
information of the interaction between the bipartite states and the
environment. (\ref{interaction 02}) is a four-point function, and in the
special case when $\lvert \phi _{0}\rangle $ is the empty vacuum state, is a
three-point function. we view $\lvert \Delta _{n}\rangle $ as giant graviton
states and $\lvert \phi _{i}\rangle $ as close string states. In the above,
\begin{equation}
U^{BC}=\sum_{i,j,n^{\prime },n}A_{n^{\prime },n}^{i,j}\lvert \Delta
_{n^{\prime }}\rangle \lvert \phi _{i}\rangle \langle \phi _{j}|\langle
\Delta _{n}|,~~\ \ \ ~A_{n^{\prime },n}^{i}=A_{n^{\prime },n}^{i,0}.
\end{equation}%
The interaction details are similar to Sec 4.3.

Tracing $C$, the superoperator $\mathcal{E}$ is
\begin{equation}
\rho _{AB}=\mathcal{E}(\rho _{AB}^{0})=\text{\textrm{tr}}_{C}~(I\otimes
U^{BC})(\rho _{AB}^{0}\otimes \rho _{C}^{0})(I\otimes U^{BC})^{\dagger },
\end{equation}%
where $\rho _{C}^{0}=\lvert \phi _{0}\rangle \langle \phi _{0}|$. We have
that%
\begin{equation}
\rho _{AB}=\sum_{n,n^{\prime },l,l^{\prime }}\sum_{i}\sqrt{p_{n}p_{l}}%
A_{n^{\prime },n}^{i}A_{l,l^{\prime }}^{i~\ast }\lvert \Delta _{n}\rangle
\lvert \Delta _{n^{\prime }}\rangle \langle \Delta _{l^{\prime }}|\langle
\Delta _{l}|.  \label{rho_ab_01}
\end{equation}%
It can also be written as $\rho _{AB}=\sum_{i}E_{i}\rho
_{AB}^{0}E_{i}^{\dagger }$ and the Kraus operator is $E_{i}=I\otimes A^{i}$,
where $A^{i}=\sum_{n^{\prime },n}A_{n^{\prime },n}^{i}\lvert \Delta
_{n^{\prime }}\rangle \langle \Delta _{n}|$ and$~\sum_{i}E_{i}^{\dagger
}E_{i}=I$. This is equivalent to $\sum_{n^{\prime },i}A_{n,n^{\prime
}}^{i~\ast }A_{n^{\prime },l}^{i}=\delta _{n,l}$. These are also equivalent
to that $U^{BC}$ is unitary.

Tracing $BC$, $\rho _{A}=\sum_{i}(\rho _{A}^{0})^{1/2}A^{i\dagger
}A^{i}(\rho _{A}^{0})^{1/2}=\rho _{A}^{0}$, and hence $S(\rho _{A})=S(\rho
_{A}^{0})$. This is because that the interaction does not involve $A$.
Tracing $AC$,
\begin{equation}
\rho _{B}=\sum_{n^{\prime },l^{\prime }}G_{n^{\prime },l^{\prime }}\lvert
\Delta _{n^{\prime }}\rangle \langle \Delta _{l^{\prime
}}|~=\sum_{i}A^{i}\rho _{B}^{0}A^{i\dagger },  \label{rho_B}
\end{equation}%
where the matrix elements of $G$ is%
\begin{equation}
G_{n^{\prime },l^{\prime }}=\langle \Delta _{n^{\prime }}|\rho _{B}\lvert
\Delta _{l^{\prime }}\rangle =\sum_{i,n}p_{n}A_{n^{\prime
},n}^{i}A_{n,l^{\prime }}^{i~\ast }.  \label{G}
\end{equation}%
Define $\rho _{B}=\mathcal{E}^{B}(\rho _{B}^{0})=\sum_{i}A^{i}\rho
_{B}^{0}A^{i\dagger }$. Hence $\rho _{AB}=\mathcal{E}(\rho
_{AB}^{0})=(I\otimes \mathcal{E}^{B})(\rho _{AB}^{0})$. Tracing $A$,%
\begin{eqnarray}
\rho _{BC} &=&\sum_{n^{\prime },i,l^{\prime },j}\sum_{n}p_{n}A_{n^{\prime
},n}^{i}A_{n,l^{\prime }}^{j~\ast }\lvert \Delta _{n^{\prime }}\rangle
\lvert \phi _{i}\rangle \langle \phi _{j}|\langle \Delta _{l^{\prime }}|
\notag \\
&=&\sum_{i,j}(A^{i}\rho _{B}^{0}A^{j\dagger })\lvert \phi _{i}\rangle
\langle \phi _{j}|.
\end{eqnarray}%
Tracing $AB$,$~\rho _{C}=\sum_{i,j}M_{ij}\lvert \phi _{i}\rangle \langle
\phi _{j}|$, where the matrix elements of $M$ is%
\begin{equation}
M_{ij}=\langle \phi _{i}|\rho _{C}\lvert \phi _{j}\rangle =\sum_{n,n^{\prime
}}p_{n}A_{n^{\prime },n}^{i}A_{n,n^{\prime }}^{j~\ast }=\mathrm{tr}(\rho
_{B}^{0}A^{j\dagger }A^{i}).~  \label{M}
\end{equation}

The mutual information between $A$ and $B$ is
\begin{equation}
I(A,B)=-\sum_{n}p_{n}\log p_{n}-\mathrm{tr}(G\log G)+\mathrm{tr}(M\log M).
\end{equation}%
The change of the mutual information between $A$ and $B$ is%
\begin{equation}
I(A,B)-I(A,B)^{0}=\sum_{n}p_{n}\log p_{n}-\mathrm{tr}(G\log G)+\mathrm{tr}%
(M\log M).  \label{mutual information 04}
\end{equation}%
Since the right side of (\ref{mutual information 04}) is also equal to $%
-I(A,C)$, and $I(A,C)=S(\rho _{A})+S(\rho _{A})-S(\rho _{AC})\geq 0$ which
is always non-negative, we have that $I(A,B)-I(A,B)^{0}\leq 0$ holds. The
mutual information between the entangled pair is non-increasing after the
interaction with the environment.

On the other hand,
\begin{equation}
I(A,B)-I(A,C)=2I^{coh}=S(\rho _{B})-S(\rho _{AB})
\end{equation}%
where $I^{coh}$ is coherent information. The coherent information has
deduced the amount of information transmitted to or leaked to the
environment \cite{Schumacher:1996dy,Horodecki:2009zz}. The detailed methods
for computing the coherent information are in \cite{Barnum
etal,Schumacher:1996dy,Horodecki:2009zz}. We also have that $%
I(A,B)+I(A,C)=2S(\rho _{A}^{0})=$ $-2\sum_{n}p_{n}\log p_{n}$ is conserved.

The entropy increase is%
\begin{equation}
\Delta S_{AB}=S(\rho _{AB})-S(\rho _{AB}^{0})=-\mathrm{tr}\text{\textrm{(}}%
M\log M),~
\end{equation}%
where $M$ is (\ref{M}). Starting from a pure state $\rho _{AB}^{0}$, if
there is no interaction of $AB$ with the environment, the entropy increase
after the evolution would be zero. Hence, the entropy increase $S(\rho
_{AB})-S(\rho _{AB}^{0})$ is related to the amount of interaction between $%
AB $ and the environment. This is the entropy production of the system,
after the interaction of the system with the environment.

There are two important regimes. In the regime that $S(\rho _{C})=S(\rho
_{AB})$ is small, the interaction of $AB$ with the environment is small, and
the coherent information is relatively bigger. On the other hand, in the
regime that $S(\rho _{C})=S(\rho _{AB})$ is big, the interaction of $AB$
with environment is big, and the coherent information is relatively smaller.

In the more general case, interactions are between $AB$ and $C$. Due to the
interaction of $AB$ with environment,
\begin{eqnarray}
\lvert \Psi _{ABC}\rangle &=&U^{ABC}\sum_{n}\sqrt{p_{n}}\lvert \Delta
_{n}\rangle \lvert \Delta _{n}\rangle \lvert \phi _{0}\rangle  \notag \\
&=&\sum_{n,n^{\prime },n^{\prime \prime },i}\sqrt{p_{n}}A_{n^{\prime \prime
},n^{\prime },n}^{i}\lvert \Delta _{n^{\prime \prime }}\rangle \lvert \Delta
_{n^{\prime }}\rangle \lvert \phi _{i}\rangle .
\end{eqnarray}%
Here,%
\begin{equation}
A_{n^{\prime \prime },n^{\prime },n}^{i}=\langle \phi _{i}|\langle \Delta
_{n^{\prime }}|\langle \Delta _{n^{^{\prime \prime }}}|U^{ABC}\lvert \Delta
_{n}\rangle \lvert \Delta _{n}\rangle \lvert \phi _{0}\rangle .
\end{equation}%
This involves six-point functions which may be decomposable in terms of
products of lower-point functions, in the context of dual conformal field
theory.

Tracing $C$, the superoperator is
\begin{equation}
\rho _{AB}=\mathcal{E}(\rho _{AB}^{0})=\text{\textrm{tr}}_{C}~(U^{ABC})(\rho
_{AB}^{0}\otimes \rho _{C}^{0})(U^{ABC})^{\dagger }.
\end{equation}%
Here$~\rho _{C}^{0}=\lvert \phi _{0}\rangle \langle \phi _{0}|$.
\begin{equation}
\rho _{AB}=\sum_{n^{\prime },n^{\prime \prime },l^{\prime },l^{\prime \prime
}}\sum_{i,n,l}\sqrt{p_{n}p_{l}}A_{n^{\prime \prime },n^{\prime
},n}^{i}A_{l,l^{\prime },l^{\prime \prime }}^{i~\ast }\lvert \Delta
_{n^{\prime \prime }}\rangle \lvert \Delta _{n^{\prime }}\rangle \langle
\Delta _{l^{\prime }}|\langle \Delta _{l^{\prime \prime }}|.
\end{equation}

Under the special case $U^{ABC}=I\otimes U^{BC}$, the scattering processes
in $A$ are completely disconnected with the scattering processes in $BC$,
and hence $A_{n^{\prime \prime },n^{\prime },n}^{i}=A_{n^{\prime
},n}^{i}\delta _{n^{\prime \prime },n}$, this recovers (\ref{interaction 02}%
).

\section{Discussion}

\label{sec_discussion}

In this paper, we used entangled multimode coherent states to produce
entangled giant graviton states. This multimode coherent state has a gravity
dual in terms of a bump on the five-sphere. We make a smeared distribution
of the entangled multimode coherent states on the circle, or on the
five-sphere, in the higher dimensional view. By the linear transformation
from multimode coherent states to giant graviton states, these states become
entangled giant graviton states. The distribution functions play the role of
the deformation of the circular edge of the droplets and are closely related
to the chiral field describing the edges of the droplets. In the context of
gauge/gravity duality, we analyzed the superposition of giant graviton
states, and the entangled pairs of giant graviton states.

We define regularized YT states and make use of them to define regularized
version of multimode coherent states. This regularization makes the Hilbert
spaces of these subsystems finite dimensional and hence simpler for
computations. These states can then become discrete-variable quantum states,
for example, the cat qudits. We use the regularized YT states as qudits.
There are many unitary quantum operations acting on them as we analyzed in
Sec 2.3 and Sec 3. The entangled giant graviton states can also be described
by Choi states, parameterized by unitary quantum operations, in which the
unitary operations are mapped from the angular distribution functions in
Prop. 2.2. The micro-macro entangled states between YT states and trace
states are also produced in Sec. 3. We analyzed various quantum operations,
such as beam splitters, squeezers and displacers, on bipartite states
involving the YT states. We have seen that the particle subtraction and
particle addition operations are very useful in increasing the entanglement
in bipartite states. In the context of gauge/gravity duality, the YT states
describe the giant graviton states.

We then produced mixed states of YT states and computed their entropies.
They can be produced by ensemble mixing of pure YT states. We use angular
distribution functions, and make a smeared distribution of giant gravitons
on the circle, or on the five-sphere, in the higher dimensional view, to
produce the mixed states. We also use noisy quantum channels, in which, by
partial tracing after the interaction with the environment, to produce the
noisy YT states. We also use noisy quantum channels to produce mixed
entangled YT states, from pure entangled YT states going through the
channel. We analyzed observables of quantum information, such as the
entropies of the subsystems and the mutual informations between the system
and environment, between the subsystems, and between the mixed entangled
pair.

The gauge/gravity duality enables us to analyze the aspects of superposition
and entanglement for the quantum gravity side. The ideas of superposition of
states on the gravity side, in the nonperturbative and global sense, have
also been considered in \cite%
{Berenstein:2017abm,Berenstein:2016pcx,Berenstein:2016mxt,Anastopoulos:2015zta,Nomura:2016aww,Lin:2017dnz}%
. The gravitational aspects of the gravitational superposition states have
been discussed in \cite%
{Berenstein:2017abm,Berenstein:2016pcx,Berenstein:2016mxt,Anastopoulos:2015zta}%
. The cat states in four dimensional gravity have been discussed in \cite%
{Anastopoulos:2015zta}. It is useful to explore these ideas with the setup
of this paper. These geometries are very explicit and they serve as a good
laboratory to perform quantitative computations. Moreover, the system is
ultraviolet finite, since it has the ultraviolet completion in string theory.

Our results may provide further insights into other interesting related
phenomena in gauge/gravity correspondence. Various other similar spacetime
geometries, in the context of string theory and quantum gravity have been
analyzed, see for example \cite%
{Mathur:2005ai,Balasubramanian:2007bs,Fareghbal:2008ar} and their related
discussion. Our methods and discussions are also related to fuzzball
proposal \cite{Mathur:2005ai}--\cite{Bena:2018mpb}. There are also
scrambling behaviors in heavy and excited states in the gauge theory duals
\cite{deMelloKoch:2020jmf,McLoughlin:2020zew}. On the other hand, scrambling
behaviors have also been observed in fuzzball geometries \cite{Bena:2018mpb}.

The approach of correlation and entanglement in phase space \cite{Almeida}
is convenient for studying questions with gravitational degrees of freedom
using different regions of the phase space \cite{Simon:2018laf,Lin:2020qao}.
The setup here naturally includes the phase space in the gravitational
system, and provides a laboratory for studying these quantum gravitational
questions. It is convenient to analyze the correlation and entanglement
between gravitational degrees of freedom using different regions of the
phase space plane in bubbling AdS. Inspired by previous works \cite%
{Berenstein:2017abm,Lin:2017dnz,Simon:2018laf}, it is very interesting to
further study the relation between entanglement and the dual spacetime
physics.

Entanglement between different parts of internal five-spheres were also
discussed in \cite{Simon:2018laf,Mollabashi:2014qfa,
Lin:2020qao,Berenstein:2017abm,Berenstein:2018lrm,Balasubramanian:2017hgy}.
In addition to different parts of the coordinate five-sphere, \cite%
{Simon:2018laf,Lin:2020qao} also used the different parts of phase space
\cite{Lin:2004nb,Almeida} to describe the entanglement. The internal
five-sphere can be generalized to more general five-manifolds and the
corresponding state space is similar, e.g. \cite{Grant:2007ze}. Entanglement
between different parts of the internal dimensions or extra dimensions, and
from string and brane degrees of freedom in spacetime, have also been
considered in \cite{Das:2020xoa,Hampapura:2020hfg,Karch:2014pma}. These
important insights are closely related to our scenarios. It would be
interesting to see more detailed relations between these insights and the
discussions here.

The approach here is interesting for understanding the emergence of
spacetime geometries, see for example \cite%
{Rangamani:2016dms,VanRaamsdonk:2010pw,Horowitz:2006ct,Koch:2009gq}. These
geometries are very explicit and they serve as a good laboratory to perform
quantitative calculations and predictions. It would also be good to
understand in more detail the relation to the scenarios of building-up
spacetime geometries, as proposed in for example \cite%
{VanRaamsdonk:2010pw,Maldacena:2013xja,Rangamani:2016dms}. Viewing the
spacetime as a quantum error correction code, is a very remarkable insight
\cite{Almheiri:2014lwa,Pastawski:2015qua,Almheiri:2016blp,Berenstein:2017rrx}
for emergent spacetime, and see also, related insights in \cite%
{Harlow:2013tf,Verlinde:2012cy}. It would be good to understand these
related aspects better, in the context of emergent spacetime and
gauge/gravity duality.

\section*{Acknowledgments}

We would like to thank B.~Czech, R.~de Mello Koch and J. Simon for
communications or discussions. The work was supported in part by Yau
Mathematical Sciences Center and Tsinghua University, by grant TH-533310008
of Tsinghua University (to H.L.), and by National Key R\&D Program of China
grant No. 2020YFA0713000.

\appendix

\section{Multimode coherent states and Young tableau states}

\renewcommand{\theequation}{A.\arabic{equation}} \setcounter{equation}{0}

\label{appendix_inner product}

In this appendix, we overview a class of single mode and multi mode coherent
states. This class was constructed in \cite{Berenstein:2017abm}, and
analyzed in further details in \cite{Lin:2017dnz,Lin:2020qao}.

Consider the Hilbert space factorizes as $\mathcal{H}$ $=\mathcal{H}%
_{1}\otimes \mathcal{H}_{2}\otimes \dots =\otimes _{k}\mathcal{H}_{k}$. Here
$\mathcal{H}_{k}$ is the Hilbert space for mode $k\in \mathbb{Z}_{>0}$. The
creation and annihilation operators for mode $k$ are $a_{k}^{\dagger }$ and $%
a_{k}$. The state in mode $k$, with occupation number $l$, is
\begin{equation}
t_{k}^{l}=(a_{k}^{\dagger })^{l}|0\rangle _{k},
\end{equation}%
where $|0\rangle _{k}~$is the vacuum of $\mathcal{H}_{k}$ and $%
t_{k}^{0}=|0\rangle _{k}$. In our convention, the normalization is $\frac{1}{%
k}[a_{k},a_{k}^{\dagger }]=1$.\ The state $\frac{1}{\sqrt{l!k^{l}}}%
(a_{k}^{\dagger })^{l}|0\rangle _{k}$ has unit norm.

This construction works generally for systems having a similar Hilbert
space, and in the context of the gauge theory in which gauge invariant
observables can be constructed from a complex matrix $Y$, the $t_{k}$
corresponds to $\mathrm{Tr~}Y^{k}$. In that case,%
\begin{equation}
a_{k}^{\dagger }|0\rangle _{k}=\frac{1}{\sqrt{N^{k}}}\text{\textrm{Tr~}}%
Y^{k},
\end{equation}%
and $t_{k}^{l}=(a_{k}^{\dagger })^{l}|0\rangle _{k}=(\frac{1}{\sqrt{N^{k}}}$%
\textrm{Tr}$(Y^{k}))^{l}$.~In the context of gauge theory, the prefactors
involving $N$ can be calculated by gauge theory computations, e.g. \cite%
{Kristjansen:2002bb,Corley:2002mj,Koch:2008cm,Berenstein:2017abm}. Here, we
work in the large $N$ limit of gauge theory, because in this limit, the
multi traces provide good orthogonality property. In the context of
gauge/string duality, $t_{k}$ is also a closed string state.

We consider coherent states generalized from the coherent states of photons
\cite{Zhang:1990fy}. A general multi-mode coherent state can be written as
\cite{Berenstein:2017abm},
\begin{equation}
|Coh(\{\Lambda _{k}\})\rangle =\prod_{k}\exp (\Lambda _{k}\frac{%
a_{k}^{\dagger }}{k})|0\rangle _{k}
\end{equation}%
where $\Lambda _{k}\in \mathbb{C}$. The parameters $\{\Lambda _{k}\}$ in $%
Coh(\{\Lambda _{k}\})~$is a family of complex parameters for the modes $k\in
\mathbb{Z}_{>0}$. These coherent states are at the level of multi-mode
coherent states in the Hilbert space $\otimes _{k}\mathcal{H}_{k}$.

These are pure coherent states. They are the eigenstates of the annihilation
operator $a_{k}$ with eigenvalue $\Lambda _{k},$%
\begin{equation}
a_{k}|Coh(\{\Lambda _{k}\})\rangle =\Lambda _{k}|Coh(\{\Lambda
_{k}\})\rangle .  \label{coherent state condition}
\end{equation}

The single-mode coherent state is a special case. We denote $|Coh(\{\Lambda
_{k}\})\rangle $ for the multi-mode one, and $|coh(\Lambda _{k})\rangle
_{k}=\exp (\Lambda _{k}\frac{a_{k}^{\dagger }}{k})|0\rangle _{k}$ for the
single-mode one in mode $k$. The subscript $k$ denotes that the state is in
mode-$k$ subspace $\mathcal{H}_{k}$. A special multi-mode\emph{\ }coherent
state is when $\Lambda _{k}:=(\Lambda )^{k}$, in which $\Lambda \in \mathbb{C%
}$. The amplitude in each mode is correlated, since $\Lambda _{k}=\Lambda
^{k}$. This multi-mode state is
\begin{equation}
|Coh(\Lambda )\rangle =\prod_{k=1}^{\infty }\exp (\Lambda ^{k}\frac{%
a_{k}^{\dagger }}{k})|0\rangle _{k}.
\end{equation}%
The normalization for $|Coh(\Lambda )\rangle ~$is%
\begin{equation}
\mathcal{N}(\Lambda )=\langle Coh(\Lambda )|Coh(\Lambda )\rangle =\exp
(\sum\limits_{k=1}^{\infty }\frac{|\Lambda |^{2k}}{k})=\frac{1}{(1-|\Lambda
|^{2})}.
\end{equation}%
Hence the inner product of two unit-norm multi-mode coherent states is
\begin{eqnarray}
&&\frac{1}{\sqrt{\mathcal{N}(\Lambda _{(0)})\mathcal{N}(\Lambda _{(1)})}}%
\langle Coh(\Lambda _{(0)})|Coh(\Lambda _{(1)})\rangle   \notag \\
&=&\frac{\left( 1-|\Lambda _{(0)}|^{2}\right) ^{1/2}\left( 1-|\Lambda
_{(1)}|^{2}\right) ^{1/2}}{(1-\bar{\Lambda}_{(0)}\Lambda _{(1)})}.
\end{eqnarray}

In addition to the single-parameter multimode coherent states, there are
also multi-parameter multimode coherent states constructed in \cite%
{Berenstein:2017abm}, and further analyzed in \cite{Lin:2017dnz,Lin:2020qao}.

For a Young tableau $\lambda $, we write $\lambda \vdash n$ to mean that $%
\lambda $ corresponds to a partition of $n$. From the representation theory
of symmetric group, we know that each Young tableau $\lambda $ is associated
with an irreducible representation of the symmetric group $S_{n}$. We define
a Young tableau state that is associated with the Young tableau $\lambda $ by%
\begin{equation}
\left\vert \lambda \right\rangle =\sum_{\vec{w}\in p(n)}\chi _{\lambda }(%
\vec{w})\prod_{k}\frac{1}{k^{w_{k}}w_{k}!}(t_{k})^{w_{k}},  \label{Def_Young}
\end{equation}%
where $p(n)$ is the set of all partitions of $n$, which is all such $\vec{w}$
that $\sum kw_{k}=w_{1}+2w_{2}+3w_{3}+\dots =n$. Here $\vec{w}$ denotes a
partition and also a conjugacy class \cite{Sagan,Fulton,Fulton Harris}. Here
$\chi _{\lambda }$ is the character of the irreducible representation
associated with $\lambda $, and $\chi _{\lambda }(\vec{w})$ means the value
of the character on the conjugacy class $\vec{w}$, or $\chi _{\lambda
}(\prod_{k}t_{k}^{w_{k}})$. The norm is $||\lambda ||=1$. These Young
tableau states were constructed in \cite{Corley:2001zk} and analyzed in
details in \cite{Berenstein:2004kk,Berenstein:2017abm,Koch:2008cm}. We can
view a general Young tableau as a multipartite system \cite%
{Berenstein:2017abm,Lin:2017dnz}, as it can be expanded by different
conjugacy classes of cycles with various lengths. They can also be expanded
as linear combinations of the multi-traces.

\end{document}